\documentclass{pasj02}
\usepackage[switch,mathlines]{lineno}
\usepackage{url}

\Received{$\langle$reception date$\rangle$}
\Accepted{$\langle$acception date$\rangle$}
\Published{$\langle$publication date$\rangle$}

\begin{document}

\title{Constraining gas motion and non-thermal pressure beyond the core of the Abell 2029 galaxy cluster with XRISM}

\author{
XRISM \textsc{Collaboration},
Marc \textsc{Audard},\altaffilmark{1} 
Hisamitsu \textsc{Awaki},\altaffilmark{2} 
Ralf \textsc{Ballhausen},\altaffilmark{3,4,5} \orcid{0000-0002-1118-8470}
Aya \textsc{Bamba},\altaffilmark{6} \orcid{0000-0003-0890-4920}
Ehud \textsc{Behar},\altaffilmark{7} \orcid{0000-0001-9735-4873}
Rozenn \textsc{Boissay-Malaquin},\altaffilmark{8,4,5} \orcid{0000-0003-2704-599X}
Laura \textsc{Brenneman},\altaffilmark{9} \orcid{0000-0003-2663-1954}
Gregory \textsc{Brown},\altaffilmark{10} \orcid{0000-0001-6338-9445}
Lia \textsc{Corrales},\altaffilmark{11} \orcid{0000-0002-5466-3817}
Elisa \textsc{Costantini},\altaffilmark{12} \orcid{0000-0001-8470-749X}
Renata \textsc{Cumbee},\altaffilmark{4} \orcid{0000-0001-9894-295X}
Maria \textsc{Diaz Trigo},\altaffilmark{13} \orcid{0000-0001-7796-4279}
Chris \textsc{Done},\altaffilmark{14} \orcid{0000-0002-1065-7239}
Tadayasu \textsc{Dotani},\altaffilmark{15} 
Ken \textsc{Ebisawa},\altaffilmark{15} \orcid{0000-0002-5352-7178}
Megan \textsc{Eckart},\altaffilmark{10} \orcid{0000-0003-3894-5889}
Dominique \textsc{Eckert},\altaffilmark{1} \orcid{0000-0001-7917-3892}
Satoshi \textsc{Eguchi},\altaffilmark{16} \orcid{0000-0003-2814-9336}
Teruaki \textsc{Enoto},\altaffilmark{17} \orcid{0000-0003-1244-3100}
Yuichiro \textsc{Ezoe},\altaffilmark{18} 
Adam \textsc{Foster},\altaffilmark{9} \orcid{0000-0003-3462-8886}
Ryuichi \textsc{Fujimoto},\altaffilmark{15} \orcid{0000-0002-2374-7073}
Yutaka \textsc{Fujita},\altaffilmark{18} \orcid{0000-0003-0058-9719}
Yasushi \textsc{Fukazawa},\altaffilmark{19} \orcid{0000-0002-0921-8837}
Kotaro \textsc{Fukushima},\altaffilmark{15} \orcid{0000-0001-8055-7113}
Akihiro \textsc{Furuzawa},\altaffilmark{20} 
Luigi \textsc{Gallo},\altaffilmark{21} \orcid{0009-0006-4968-7108}
Javier \textsc{Garc\'ia},\altaffilmark{4,22} \orcid{0000-0003-3828-2448}
Liyi \textsc{Gu},\altaffilmark{12} \orcid{0000-0001-9911-7038}
Matteo \textsc{Guainazzi},\altaffilmark{23} \orcid{0000-0002-1094-3147}
Kouichi \textsc{Hagino},\altaffilmark{6} \orcid{0000-0003-4235-5304}
Kenji \textsc{Hamaguchi},\altaffilmark{8,4,5} \orcid{0000-0001-7515-2779}
Isamu \textsc{Hatsukade},\altaffilmark{24} \orcid{0000-0003-3518-3049}
Katsuhiro \textsc{Hayashi},\altaffilmark{15} \orcid{0000-0001-6922-6583}
Takayuki \textsc{Hayashi},\altaffilmark{8,4,5} \orcid{0000-0001-6665-2499}
Natalie \textsc{Hell},\altaffilmark{10} \orcid{0000-0003-3057-1536}
Edmund \textsc{Hodges-Kluck},\altaffilmark{4}\orcid{0000-0002-2397-206X}
Ann \textsc{Hornschemeier},\altaffilmark{4} \orcid{0000-0001-8667-2681}
Yuto \textsc{Ichinohe},\altaffilmark{25} \orcid{0000-0002-6102-1441}
Daiki \textsc{Ishi},\altaffilmark{15}
Manabu \textsc{Ishida},\altaffilmark{15} 
Kumi \textsc{Ishikawa},\altaffilmark{18} 
Yoshitaka \textsc{Ishisaki},\altaffilmark{18} 
Jelle \textsc{Kaastra},\altaffilmark{12,26} \orcid{0000-0001-5540-2822}
Timothy \textsc{Kallman},\altaffilmark{4} 
Erin \textsc{Kara},\altaffilmark{27} \orcid{0000-0003-0172-0854}
Satoru \textsc{Katsuda},\altaffilmark{28} \orcid{0000-0002-1104-7205}
Yoshiaki \textsc{Kanemaru},\altaffilmark{15} \orcid{0000-0002-4541-1044}
Richard \textsc{Kelley},\altaffilmark{4} \orcid{0009-0007-2283-3336}
Caroline \textsc{Kilbourne},\altaffilmark{4} \orcid{0000-0001-9464-4103}
Shunji \textsc{Kitamoto},\altaffilmark{29} \orcid{0000-0001-8948-7983}
Shogo \textsc{Kobayashi},\altaffilmark{30} \orcid{0000-0001-7773-9266}
Takayoshi \textsc{Kohmura},\altaffilmark{31} 
Aya \textsc{Kubota},\altaffilmark{32} 
Maurice \textsc{Leutenegger},\altaffilmark{4} \orcid{0000-0002-3331-7595}
Michael \textsc{Loewenstein},\altaffilmark{3,4,5} \orcid{0000-0002-1661-4029}
Yoshitomo \textsc{Maeda},\altaffilmark{15} \orcid{0000-0002-9099-5755}
Maxim \textsc{Markevitch},\altaffilmark{4} 
Hironori \textsc{Matsumoto},\altaffilmark{33} 
Kyoko \textsc{Matsushita},\altaffilmark{30} \orcid{0000-0003-2907-0902}
Dan \textsc{McCammon},\altaffilmark{34} \orcid{0000-0001-5170-4567}
Brian \textsc{McNamara},\altaffilmark{35} 
Francois \textsc{Mernier},\altaffilmark{3,4,5} \orcid{0000-0002-7031-4772}
Eric \textsc{Miller},\altaffilmark{27} \orcid{0000-0002-3031-2326}
Jon \textsc{Miller},\altaffilmark{11} \orcid{0000-0003-2869-7682}
Ikuyuki \textsc{Mitsuishi},,\altaffilmark{36}\orcid{0000-0002-9901-233X}
Misaki \textsc{Mizumoto},\altaffilmark{37} \orcid{0000-0003-2161-0361}
Tsunefumi \textsc{Mizuno},\altaffilmark{38} \orcid{0000-0001-7263-0296}
Koji \textsc{Mori},\altaffilmark{24} \orcid{0000-0002-0018-0369}
Koji \textsc{Mukai},\altaffilmark{8,4,5} \orcid{0000-0002-8286-8094}
Hiroshi \textsc{Murakami},\altaffilmark{39} 
Richard \textsc{Mushotzky},\altaffilmark{3} \orcid{0000-0002-7962-5446}
Hiroshi \textsc{Nakajima},\altaffilmark{40} \orcid{0000-0001-6988-3938}
Kazuhiro \textsc{Nakazawa},\altaffilmark{36} \orcid{0000-0003-2930-350X}
Jan-Uwe \textsc{Ness},\altaffilmark{41} 
Kumiko \textsc{Nobukawa},\altaffilmark{42} \orcid{0000-0002-0726-7862}
Masayoshi \textsc{Nobukawa},\altaffilmark{43} \orcid{0000-0003-1130-5363}
Hirofumi \textsc{Noda},\altaffilmark{44} \orcid{0000-0001-6020-517X}
Hirokazu \textsc{Odaka},\altaffilmark{33} 
Shoji \textsc{Ogawa},\altaffilmark{15} \orcid{0000-0002-5701-0811}
Anna \textsc{Ogorzalek},\altaffilmark{3,4,5} \orcid{0000-0003-4504-2557}
Takashi \textsc{Okajima},\altaffilmark{4} \orcid{0000-0002-6054-3432}
Naomi \textsc{Ota},\altaffilmark{45}\altemailmark \orcid{0000-0002-2784-3652}\email{naomi@cc.nara-wu.ac.jp}
Stephane \textsc{Paltani},\altaffilmark{1} \orcid{0000-0002-8108-9179}
Robert \textsc{Petre},\altaffilmark{4} \orcid{0000-0003-3850-2041}
Paul \textsc{Plucinsky},\altaffilmark{9} \orcid{0000-0003-1415-5823}
Frederick \textsc{Porter},\altaffilmark{4} \orcid{0000-0002-6374-1119}
Katja \textsc{Pottschmidt},\altaffilmark{8,4,5} \orcid{0000-0002-4656-6881}
Kosuke \textsc{Sato},\altaffilmark{28,65} \orcid{0000-0001-5774-1633}
Toshiki \textsc{Sato},\altaffilmark{47} 
Makoto \textsc{Sawada},\altaffilmark{29} \orcid{0000-0003-2008-6887}
Hiromi \textsc{Seta},\altaffilmark{18} 
Megumi \textsc{Shidatsu},\altaffilmark{2} \orcid{0000-0001-8195-6546}
Aurora \textsc{Simionescu},\altaffilmark{12} \orcid{0000-0002-9714-3862}
Randall \textsc{Smith},\altaffilmark{9} \orcid{0000-0003-4284-4167}
Hiromasa \textsc{Suzuki},\altaffilmark{24} \orcid{0000-0002-8152-6172}
Andrew \textsc{Szymkowiak},\altaffilmark{48} \orcid{0000-0002-4974-687X}
Hiromitsu \textsc{Takahashi},\altaffilmark{19} \orcid{0000-0001-6314-5897}
Mai \textsc{Takeo},\altaffilmark{28} 
Toru \textsc{Tamagawa},\altaffilmark{25} 
Keisuke \textsc{Tamura},\altaffilmark{8,4,5} 
Takaaki \textsc{Tanaka},\altaffilmark{49} \orcid{0000-0002-4383-0368}
Atsushi \textsc{Tanimoto},\altaffilmark{50} \orcid{0000-0002-0114-5581}
Makoto \textsc{Tashiro},\altaffilmark{28,15} \orcid{0000-0002-5097-1257}
Yukikatsu \textsc{Terada},\altaffilmark{28,15} \orcid{0000-0002-2359-1857}
Yuichi \textsc{Terashima},\altaffilmark{2} \orcid{0000-0003-1780-5481}
Yohko \textsc{Tsuboi},\altaffilmark{51} 
Masahiro \textsc{Tsujimoto},\altaffilmark{15} \orcid{0000-0002-9184-5556}
Hiroshi \textsc{Tsunemi},\altaffilmark{33} 
Takeshi \textsc{Tsuru},\altaffilmark{17} \orcid{0000-0002-5504-4903}
Hiroyuki \textsc{Uchida},\altaffilmark{17} \orcid{0000-0003-1518-2188}
Nagomi \textsc{Uchida},\altaffilmark{15} \orcid{0000-0002-5641-745X}
Yuusuke \textsc{Uchida},\altaffilmark{31} \orcid{0000-0002-7962-4136}
Hideki \textsc{Uchiyama},\altaffilmark{52} \orcid{0000-0003-4580-4021}
Yoshihiro \textsc{Ueda},\altaffilmark{53} \orcid{0000-0001-7821-6715}
Shinichiro \textsc{Uno},\altaffilmark{54} 
Jacco \textsc{Vink},\altaffilmark{55,12} \orcid{0000-0002-4708-4219}
Shin \textsc{Watanabe},\altaffilmark{15} \orcid{0000-0003-0441-7404}
Brian J.\ \textsc{Williams},\altaffilmark{4} \orcid{0000-0003-2063-381X}
Satoshi \textsc{Yamada},\altaffilmark{56} \orcid{0000-0002-9754-3081}
Shinya \textsc{Yamada},\altaffilmark{29} \orcid{0000-0003-4808-893X}
Hiroya \textsc{Yamaguchi},\altaffilmark{15} \orcid{0000-0002-5092-6085}
Kazutaka \textsc{Yamaoka},\altaffilmark{36} \orcid{0000-0003-3841-0980}
Noriko \textsc{Yamasaki},\altaffilmark{15} \orcid{0000-0003-4885-5537}
Makoto \textsc{Yamauchi},\altaffilmark{24} \orcid{0000-0003-1100-1423}
Shigeo \textsc{Yamauchi},\altaffilmark{57} 
Tahir \textsc{Yaqoob},\altaffilmark{8,4,5} 
Tomokage \textsc{Yoneyama},\altaffilmark{51} 
Tessei \textsc{Yoshida},\altaffilmark{15} 
Mihoko \textsc{Yukita},\altaffilmark{58,4} \orcid{0000-0001-6366-3459}
Irina \textsc{Zhuravleva},\altaffilmark{59} \orcid{0000-0001-7630-8085}
Tommaso \textsc{Bartalesi},\altaffilmark{60,61} \orcid{0009-0004-5838-2213}
Stefano \textsc{Ettori},\altaffilmark{61,62} \orcid{0000-0003-4117-8617}
Roman \textsc{Kosarzycki},\altaffilmark{63,4} 
Lorenzo \textsc{Lovisari},\altaffilmark{64,9} \orcid{0000-0002-3754-2415}
Tom \textsc{Rose},\altaffilmark{35} \orcid{0000-0002-8310-2218}
Arnab \textsc{Sarkar},\altaffilmark{27} \orcid{0000-0002-5222-1337}
Ming \textsc{Sun},\altaffilmark{65} \orcid{0000-0001-5880-0703}
Prathamesh \textsc{Tamhane}\altaffilmark{65} \orcid{0000-0001-8176-7665}}
\altaffiltext{1}{Department of Astronomy, University of Geneva, Versoix CH-1290, Switzerland}
\altaffiltext{2}{Department of Physics, Ehime University, Ehime 790-8577, Japan}
\altaffiltext{3}{Department of Astronomy, University of Maryland, College Park, MD 20742, USA}
\altaffiltext{4}{NASA / Goddard Space Flight Center, Greenbelt, MD 20771, USA}
\altaffiltext{5}{Center for Research and Exploration in Space Science and Technology, NASA / GSFC (CRESST II), Greenbelt, MD 20771, USA}
\altaffiltext{6}{Department of Physics, University of Tokyo, Tokyo 113-0033, Japan}
\altaffiltext{7}{Department of Physics, Technion, Technion City, Haifa 3200003, Israel}
\altaffiltext{8}{Center for Space Sciences and Technology, University of Maryland, Baltimore County (UMBC), Baltimore, MD, 21250 USA}
\altaffiltext{9}{Center for Astrophysics | Harvard-Smithsonian, Cambridge, MA 02138, USA}
\altaffiltext{10}{Lawrence Livermore National Laboratory, Livermore, CA 94550, USA}
\altaffiltext{11}{Department of Astronomy, University of Michigan, Ann Arbor, MI 48109, USA}
\altaffiltext{12}{SRON Netherlands Institute for Space Research, Leiden, The Netherlands}
\altaffiltext{13}{ESO, Karl-Schwarzschild-Strasse 2, 85748, Garching bei M\"{u}nchen, Germany}
\altaffiltext{14}{Centre for Extragalactic Astronomy, Department of Physics, University of Durham, Durham DH1 3LE, UK}
\altaffiltext{15}{Institute of Space and Astronautical Science (ISAS), Japan Aerospace Exploration Agency (JAXA), Kanagawa 252-5210, Japan}
\altaffiltext{16}{Department of Economics, Kumamoto Gakuen University, Kumamoto 862-8680 Japan}
\altaffiltext{17}{Department of Physics, Kyoto University, Kyoto 606-8502, Japan}
\altaffiltext{18}{Department of Physics, Tokyo Metropolitan University, Tokyo 192-0397, Japan}
\altaffiltext{19}{Department of Physics, Hiroshima University, Hiroshima 739-8526, Japan}
\altaffiltext{20}{Department of Physics, Fujita Health University, Aichi 470-1192, Japan}
\altaffiltext{21}{Department of Astronomy and Physics, Saint Mary's University, Nova Scotia B3H 3C3, Canada}
\altaffiltext{22}{California Institute of Technology, Pasadena, CA 91125, USA}
\altaffiltext{23}{European Space Agency (ESA), European Space Research and Technology Centre (ESTEC), 2200 AG Noordwijk, The Netherlands}
\altaffiltext{24}{Faculty of Engineering, University of Miyazaki, 1-1 Gakuen-Kibanadai-Nishi, Miyazaki, Miyazaki 889-2192, Japan}
\altaffiltext{25}{RIKEN Nishina Center, Saitama 351-0198, Japan}
\altaffiltext{26}{Leiden Observatory, University of Leiden, P.O. Box 9513, NL-2300 RA, Leiden, The Netherlands}
\altaffiltext{27}{Kavli Institute for Astrophysics and Space Research, Massachusetts Institute of Technology, MA 02139, USA}
\altaffiltext{28}{Department of Physics, Saitama University, Saitama 338-8570, Japan}
\altaffiltext{29}{Department of Physics, Rikkyo University, Tokyo 171-8501, Japan}
\altaffiltext{30}{Faculty of Physics, Tokyo University of Science, Tokyo 162-8601, Japan}
\altaffiltext{31}{Faculty of Science and Technology, Tokyo University of Science, Chiba 278-8510, Japan}
\altaffiltext{32}{Department of Electronic Information Systems, Shibaura Institute of Technology, Saitama 337-8570, Japan}
\altaffiltext{33}{Department of Earth and Space Science, Osaka University, Osaka 560-0043, Japan}
\altaffiltext{34}{Department of Physics, University of Wisconsin, WI 53706, USA}
\altaffiltext{35}{Department of Physics \& Astronomy, Waterloo Centre for Astrophysics, University of Waterloo, Ontario N2L 3G1, Canada}
\altaffiltext{36}{Department of Physics, Nagoya University, Aichi 464-8602, Japan}
\altaffiltext{37}{Science Research Education Unit, University of Teacher Education Fukuoka, Fukuoka 811-4192, Japan}
\altaffiltext{38}{Hiroshima Astrophysical Science Center, Hiroshima University, Hiroshima 739-8526, Japan}
\altaffiltext{39}{Department of Data Science, Tohoku Gakuin University, Miyagi 984-8588}
\altaffiltext{40}{College of Science and Engineering, Kanto Gakuin University, Kanagawa 236-8501, Japan}
\altaffiltext{41}{European Space Agency(ESA), European Space Astronomy Centre (ESAC), E-28692 Madrid, Spain}
\altaffiltext{42}{Department of Science, Faculty of Science and Engineering, KINDAI University, Osaka 577-8502, JAPAN}
\altaffiltext{43}{Department of Teacher Training and School Education, Nara University of Education, Nara 630-8528, Japan}
\altaffiltext{44}{Astronomical Institute, Tohoku University, Miyagi 980-8578, Japan}
\altaffiltext{45}{Department of Physics, Nara Women's University, Nara 630-8506, Japan}
\altaffiltext{46}{International Center for Quantum-field Measurement Systems for Studies of the Universe and Particles (QUP) / High Energy Accelerator Research Organization (KEK), 1-1 Oho, Tsukuba, Ibaraki 305-0801, Japan}
\altaffiltext{47}{School of Science and Technology, Meiji University, Kanagawa, 214-8571, Japan}
\altaffiltext{48}{Yale Center for Astronomy and Astrophysics, Yale University, CT 06520-8121, USA}
\altaffiltext{49}{Department of Physics, Konan University, Hyogo 658-8501, Japan}
\altaffiltext{50}{Graduate School of Science and Engineering, Kagoshima University, Kagoshima, 890-8580, Japan}
\altaffiltext{51}{Department of Physics, Chuo University, Tokyo 112-8551, Japan}
\altaffiltext{52}{Faculty of Education, Shizuoka University, Shizuoka 422-8529, Japan}
\altaffiltext{53}{Department of Astronomy, Kyoto University, Kyoto 606-8502, Japan}
\altaffiltext{54}{Nihon Fukushi University, Shizuoka 422-8529, Japan}
\altaffiltext{55}{Anton Pannekoek Institute, the University of Amsterdam, Postbus 942491090 GE Amsterdam, The Netherlands}
\altaffiltext{56}{RIKEN Cluster for Pioneering Research, Saitama 351-0198, Japan}
\altaffiltext{57}{Department of Physics, Faculty of Science, Nara Women's University, Nara 630-8506, Japan}
\altaffiltext{58}{Johns Hopkins University, MD 21218, USA}
\altaffiltext{59}{Department of Astronomy and Astrophysics, University of Chicago, Chicago, IL 60637, USA}
\altaffiltext{60}{Dipartimento di Fisica e Astronomia “Augusto Righi” -- Alma Mater Studiorum -- Universit\`{a} di Bologna, I-40129 Bologna}
\altaffiltext{61}{INAF, Osservatorio di Astrofisica e Scienza dello Spazio, 40129 Bologna, Italy}
\altaffiltext{62}{INFN, Sezione di Bologna, 40127 Bologna, Italy}
\altaffiltext{63}{Department of Physics, The George Washington University, Washington, DC 20052, USA}
\altaffiltext{64}{INAF -- IASF Istituto di Astrofisica Spaziale e Fisica Cosmica di Milano, Via Alfonso Corti 12, 20133 Milano, Italy}
\altaffiltext{65}{Department of Physics and Astronomy, The University of Alabama in Huntsville, Huntsville, AL 35899, USA}


\maketitle

\begin{abstract}
We report a detailed spectroscopic study of the gas dynamics and hydrostatic mass bias of the galaxy cluster Abell 2029, utilizing high-resolution observations from XRISM Resolve. Abell 2029, known for its cool core and relaxed X-ray morphology, provides an excellent opportunity to investigate the influence of gas motions beyond the central region. Expanding upon prior studies that revealed low turbulence and bulk motions within the core, our analysis covers regions out to the scale radius $R_{2500}$ (670~kpc) based on three radial pointings extending from the cluster center toward the northern side. We obtain accurate measurements of bulk and turbulent velocities along the line of sight. The results indicate that non-thermal pressure accounts for no more than 2\% of the total pressure at all radii, with a gradual decrease outward. The observed radial trend differs from many numerical simulations, which often predict an increase in non-thermal pressure fraction at larger radii. These findings suggest that deviations from hydrostatic equilibrium are small, leading to a hydrostatic mass bias of around 2\% across the observed area.
\end{abstract}


\section{Introduction}
Galaxy clusters, the largest gravitationally bound systems in the universe, grow through the hierarchical assembly of matter, accumulating mass via mergers and accretion from the cosmic web. Understanding the gas dynamics in these clusters is critical for studying their energy budget, as a substantial fraction of the injected kinetic energy stirs up the intracluster gas, creating bulk motions and turbulence that can remain unthermalized for billions of years \citep{Norman99,Miniati14,Vazza09}. While the hot intracluster medium (ICM) of clusters is often assumed to be hydrostatic, both simulations \citep{Lau09,Biffi16} and observations \citep{Ota16,Sanders20,Hitomi18} have suggested that even seemingly regular systems can exhibit significant gas motions. This prompts a key question: are `relaxed' clusters genuinely in dynamical equilibrium, or does their calm appearance conceal significant bulk motions and turbulence?

The galaxy cluster Abell~2029 (A2029) is a prototypical massive galaxy cluster with a cool core and an apparently relaxed X-ray morphology \citep{Lewis02}, located at $z=0.0787$ \citep{Sohn19}. However, the unsharp-masked X-ray surface brightness image reveals a spiral 'sloshing' structure, which may indicate significant gas motion in the core region \citep{Paterno-Mahler13}. Similar features have been identified in many clusters \citep{Ueda20}, challenging the simplistic assumption of a perfectly hydrostatic system and pointing to a more complex dynamical state. Furthermore, cosmological simulations suggest that the degree of gas disturbance correlates with a cluster's connectivity to large-scale filamentary structures, highlighting the interplay between cluster-scale dynamics and cosmic-scale structure formation \citep{Zinger16, Gouin21}.

In the first paper of our observational series on A2029, utilizing the XRISM satellite \citep{Tashiro25,Tashiro20} and its Resolve detector \citep{Ishisaki22},  we focused on the inner region of A2029, aiming to directly measure the turbulent and bulk motions in the core \citep{XRISM25}. The results indicated that non-thermal pressure contributes only a small fraction ($\sim 2$\%) to the total pressure in the cluster center, consistent with a subsonic, dynamically stable core. However, the thermodynamic state of the gas in the outer regions, particularly at radii approaching the scale radius $R_{2500}$ \footnote{The radius $R_{\Delta}$ is defined as the radius within which the average mass density of the galaxy cluster is $\Delta$ times the critical density of the universe at the cluster redshift. In this paper, we specifically use $R_{2500}$ and $R_{200}$, corresponding to $\Delta=2500$ and $\Delta=200$,respectively.} remains unclear. This marks an essential unresolved issue, as the non-thermal pressure fraction at larger radii plays a significant role in determining the hydrostatic mass bias -- the systematic offset between the true mass of the cluster and that inferred assuming hydrostatic equilibrium \citep{Nelson14}. Accurately characterizing this bias is essential for deriving robust cosmological constraints from cluster mass measurements \citep{Allen08,Pratt19}.

In this paper, we extend the analysis of A2029 to cover the outer regions up to $R_{2500} \sim 670$~kpc \citep{Vikhlinin06}. By combining XRISM’s high-resolution spectral data across three radial pointings, we simultaneously constrain the bulk and turbulent velocity components of the ICM. This enables a precise estimate of the non-thermal pressure support as a function of radius. Finally, we discuss the implications of our findings for hydrostatic mass bias and compare our results with predictions from numerical simulations and other observations. Our study highlights the importance of disentangling gas motions to refine galaxy clusters' use as cosmological probes. 

The cosmological parameters are $\Omega_{m0}=0.3$, $\Omega_\Lambda=0.7$ and $h=0.7$ throughout this paper. Accordingly, $1\arcmin$ corresponds to 89~kpc at the redshift of A2029 ($z = 0.0787$). We report redshifts and velocities corrected to the Solar System barycenter. We use the proto-solar abundance table from \citet{Lodders09}. The quoted errors represent the $1\sigma$ statistical uncertainties, unless stated otherwise.

\section{Observations and data reduction}
\begin{figure}[htb]
    \begin{center}
    \includegraphics[width=0.8\linewidth]{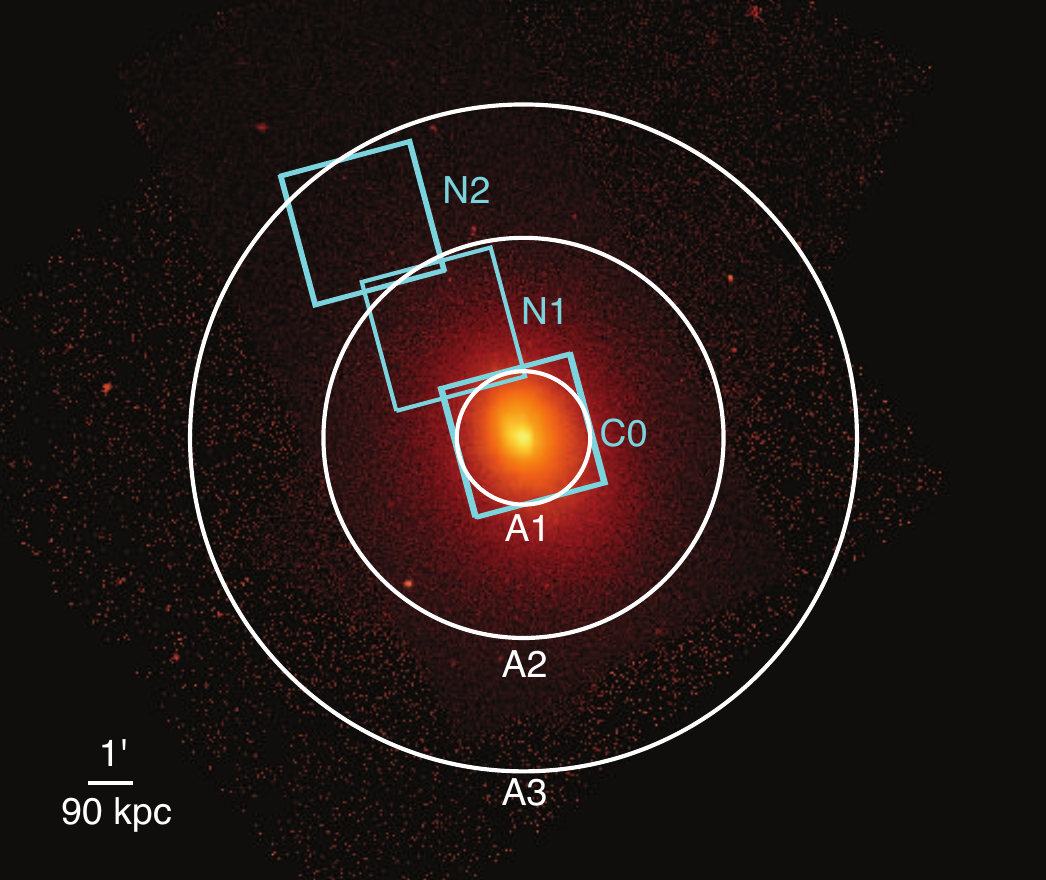}
    \end{center}
    \caption{Chandra image of A2029 overlaid with the Resolve field of view covering up to $R_{2500}$ with three radial pointings (C0, N1, N2 boxes). The three annuli (A1, A2, A3) correspond to radii of $r<1\arcmin.5$, $1\arcmin.5<r<4\arcmin.5$, and $4\arcmin.5<r<7\arcmin.5$, respectively, with the outermost circle marking $R_{2500}\sim 670$~kpc. {Alt text: Three boxes are arranged radially from the cluster center, indicating the positions of the detector's field of view.} }\label{fig:resolve_field}\end{figure}

To accurately measure the gas velocity at a radius of $R_{2500}$, it is essential to precisely separate the emission from the central cool core of the galaxy cluster, which can mix due to the telescope's point spread function (PSF) effect. For this purpose, we covered up to $R_{2500}=7\arcmin.5 \sim 670$~kpc using three consecutive radial pointings from the center (figure~\ref{fig:resolve_field}). By simultaneously fitting the spectra from these regions, we disentangled the contributions from each radius. The XRISM observations were conducted five times in January and July 2024. The observation details are given in table~\ref{tab:observation_log}.

The data reduction for the Resolve detector was performed in accordance with the XRISM Quick Start Guide v2.1. Specifically, we used Data Processing Version 03.00.011.008, reprocessed with Build8 using the standard \texttt{xapipeline} procedure. The software and calibration database (CalDB) used were XRISM\_20Jun2024\_Build8 and CalDB 8, respectively. Rise-time filtering and pixel-to-pixel coincidence screening were  applied to the pipeline-produced cleaned-events files to minimize the background. The exposure times  are summarized in Table~\ref{tab:observation_log}. A total of 549~ksec of clean data was obtained in the three pointings for the analysis.

Each spectrum was obtained by integrating the high-resolution (Hp) grade events across all pixels in the Resolve field of view, excluding pixel 27, which is omitted due to small, abrupt changes in its energy scale that cannot be corrected using the current gain-monitoring procedure. Pixel 11 also experiences such abrupt changes, but infrequently, and almost always in a correctable part of the observations. However, we identified a small jump ($\sim3$~eV at 6~keV) in the energy scale of pixel 11 during OBSID 000150000 (N1), affecting 5 ks of exposure, via comparison to the gain trend on the other pixels, after much of the analysis had been completed. To investigate the impact of this interval, we compared full-field analysis of the N1 data with and without this interval of time and found no differences exceeding the statistical errors.

Spectral fitting was performed using the Cash statistic \citep{Cash79}, which is theoretically valid even when some spectral bins contain zero observed counts. However, in practice, fits can become numerically unstable when a large number of zero-count bins are present, particularly in high-resolution spectra. To improve computational efficiency, we applied optimal binning using \texttt{ftgrouppha} \citep{Kaastra16}, requiring at least one count per bin. This allows us to reduce the number of bins while retaining essential spectral information. The impact of this binning choice is further examined in section~\ref{subsec:systematic} iv). Furthermore, due to the limited number of iron-line counts in the first N2 observation (OBSID 000152000), this dataset was excluded from the spectral analysis. Although Poisson-based statistics are applicable in principle, we found that including this observation did not improve the photon statistics or the robustness of the fits.

The ARF file, which accounts for the effective area by incorporating detector efficiencies and the X-ray Mirror Assembly (XMA) response (including PSF effects), was generated using xaarfgen with the X-ray image from the Chandra satellite as the input brightness distribution. To account for the spatial redistribution of photons due to the telescope's PSF, we performed a spatial-spectral mixing analysis: the photon observed in each spectral extraction region (C0, N1, N2) can originate from multiple spatial annuli (A1, A2, A3), as shown in figure~\ref{fig:resolve_field}. Therefore, for each OBSID, we generated three separate ARFs corresponding to the contributions from A1, A2, and A3 to each of the three spectral regions. In the spectral fitting, we constructed emission models for the A1, A2, and A3 regions, and convolved each of them with the ARF corresponding to their projected contribution to the observed region (C0, N1, or N2). The sum of these convolved models was then compared to the observed spectra. This method has been successfully demonstrated in the Hitomi satellite's Perseus cluster study \citep{Hitomi18}, and allows us to correct for PSF-induced cross-contamination between annuli.

The detector background was generated using \texttt{rslnxbgen}  on a database constructed from dark-earth eclipse data that was screened using the same criteria applied to the data, and the normalization was optimized by fitting a template model to it. Additionally, the cosmic X-ray background (CXB) component was estimated from the region outside the $R_{200}$ radius of A2029, based on the analysis of the Xtend data. For further details on the CXB estimation, refer to Sarkar et al. (in prep.). These background components were accounted for by including their optimized models as additive components in the spectral fitting process.

\begin{table*}[htb]
 \tbl{Observation Log for Abell 2029}{
    \begin{tabular}{lllllll}\hline\hline
        Object & OBSID &  Coordinates$^{a}$ & DATE  & Exposure$^{b}$ & Fe XXV He$\alpha$$^{c}$ & Fe XXVI Ly$\alpha$$^{c}$ \\
        \hline
        Abell2029\_Center &  000149000 & 227.7301, 5.7458 & 2024-01-10 & 12.4 & 294 & 115 \\
        Abell2029\_N1     &  000150000 & 227.7595, 5.7856 & 2024-01-10 & 102.2 & 289 & 129 \\
        Abell2029\_Center &  000151000 & 227.7291, 5.7457 & 2024-01-13 & 25.1 & 616 & 236 \\
        Abell2029\_N2     &  000152000 & 227.7898, 5.8256 & 2024-01-13 & 42.9 & 13  & 2   \\
        Abell2029\_N2     &  300053010 & 227.7984, 5.8239 & 2024-07-27 & 366.2 & 104 & 83  \\
        \hline
    \end{tabular}}    \label{tab:observation_log}
    \begin{tabnote}
        $^{\mathrm{a}}$ Pointing coordinates. $^{\mathrm{b}}$ Net exposure time after data screening in kilo seconds.  $^{\mathrm{c}}$ Iron line counts without continuum component.
    \end{tabnote}
\end{table*}

\section{Analysis and results}
\subsection{Spectral analysis}
The X-ray emission from each source region was assumed to originate from thermal emission of a single-temperature plasma. To account for gas turbulence and bulk motion, the BAPEC model \citep{Smith01,Foster12} in the XSPEC software version 12.14.1 \citep{Arnaud96} was employed. In this model, the gas temperature, metallicity, and normalization, along with the Gaussian sigma of velocity broadening ($\sigma_v$, in ${\rm km\,s^{-1}}$) and redshift, were treated as free parameters. The Galactic absorption was modeled using the TBabs model \citep{Wilms00}, with the hydrogen column density fixed at $3\times10^{20}~{\rm cm^{-2}}$ \citep{HI4PI16}. 

We performed the spectral fitting in the 2--10 keV band. As reported in \citet{XRISM25}, the two-temperature component in the core region was significant in the central area. However, it was confirmed that this did not affect the velocity measurements. For the N1 and N2 regions, we tested a two-temperature model but found that the additional component was not statistically significant. Therefore, the results in this paper are reported based on the single-temperature model. A detailed investigation of the multi-temperature model will be presented in Sarkar et al. (in prep.).

Barycentric corrections of $+26~{\rm km\,s^{-1}}$ were applied to the observations taken in January 2024 (OBSID 000149000, 000150000, and 000151000), which include the Center and N1 regions. A correction of $-27~{\rm km\,s^{-1}}$ was applied to the July 2024 observation of the N2 region (OBSID 300053010). The difference of $53~{\rm km\,s^{-1}}$ arises due to the approximately six-month gap between the two observing periods. 
When simultaneously fitting all three spectra, a relative barycentric correction of $-53~{\rm km\,s^{-1}}$ was applied to the cluster redshift in the N2 spectrum, aligning it with the reference frame of the January 2024 data. After the fit, an additional $+26~{\rm km\,s^{-1}}$ correction was added to derive the final redshift.

\begin{figure*}[htb]
    \begin{center}
    \includegraphics[width=0.60\linewidth]{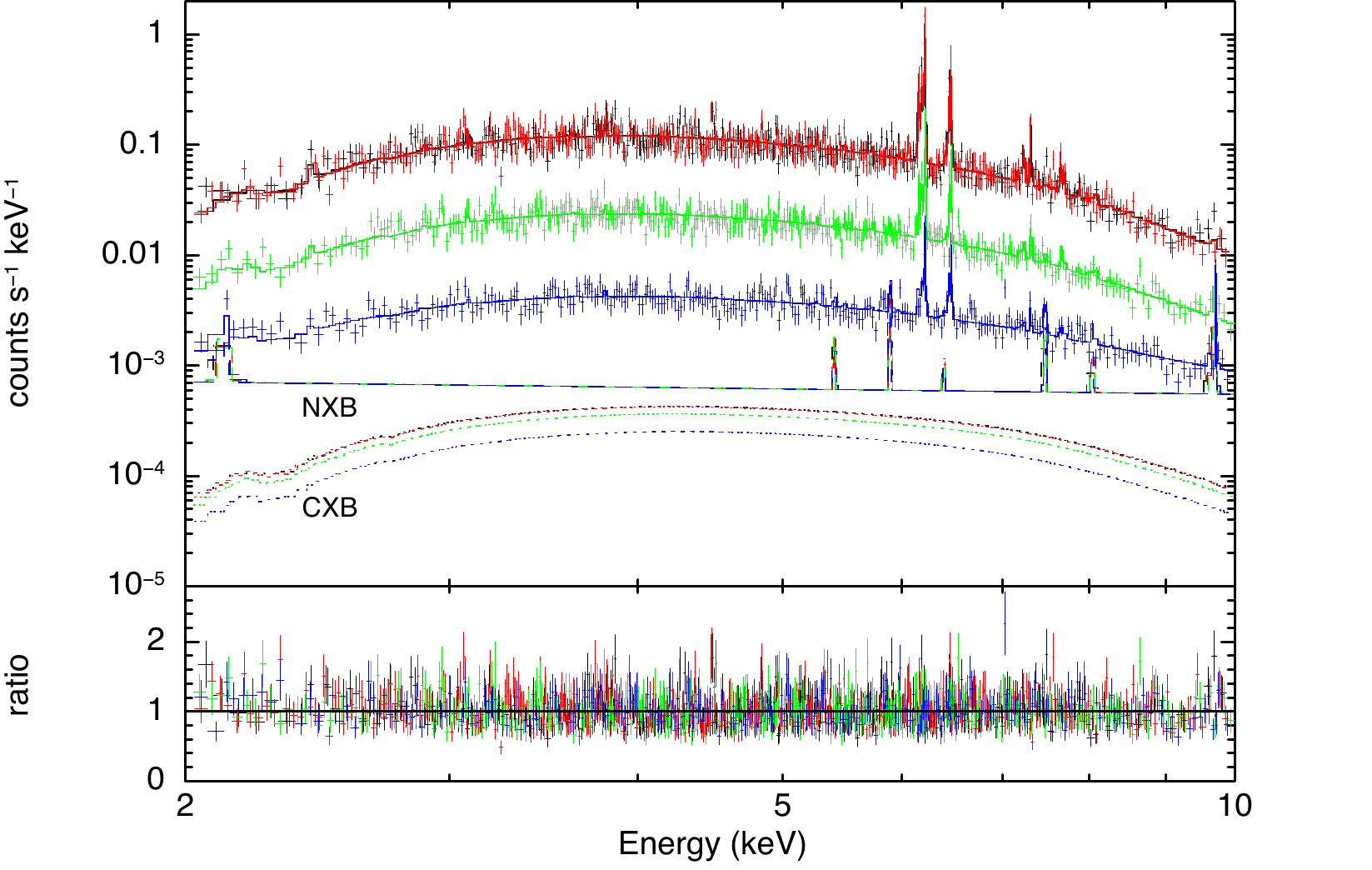}
    \end{center}
    \caption{Spectral fitting of three pointing regions in the 2-10 keV band: Center (black and red), N1 (green), and N2 (blue). The observed spectra (data points with error bars) include both the source and background components. Solid lines indicate the best-fit BAPEC models for each region; dashed lines show the non-X-ray background (NXB), and dotted lines represent the cosmic X-ray background (CXB). The spectra are binned to ensure at least 25 counts per bin for better visibility. The lower panels show the data-to-model ratio. {Alt text: Line graph showing spectra from 2 to 10 kilo electron volt in units of counts per second per kilo electron volt and their fitted curves. }}
    \label{fig:spectral_fitting}
\end{figure*}

\begin{figure*}[hbt]
    \begin{center}
    \includegraphics[width=0.45\linewidth]{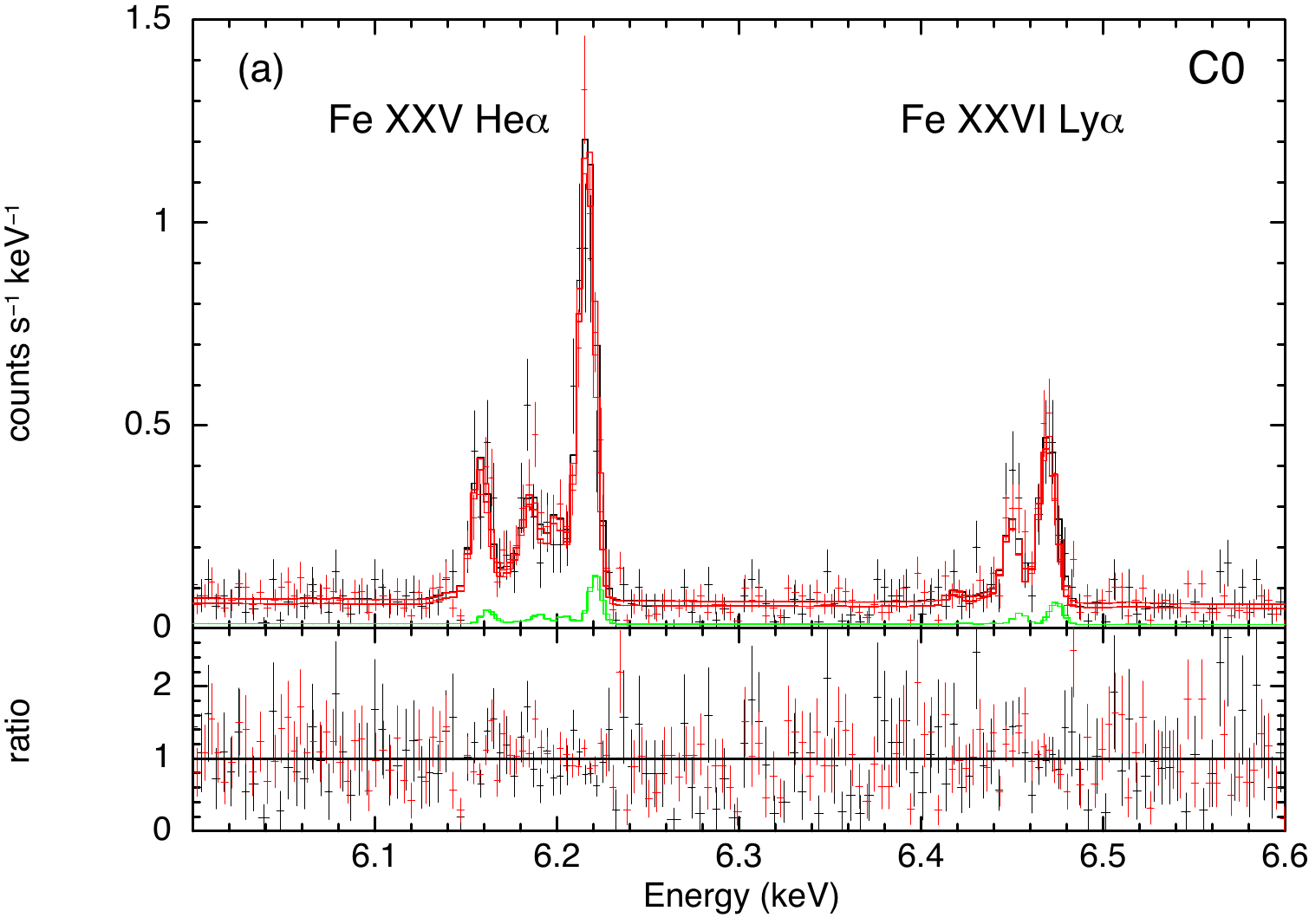}
    \includegraphics[width=0.45\linewidth]{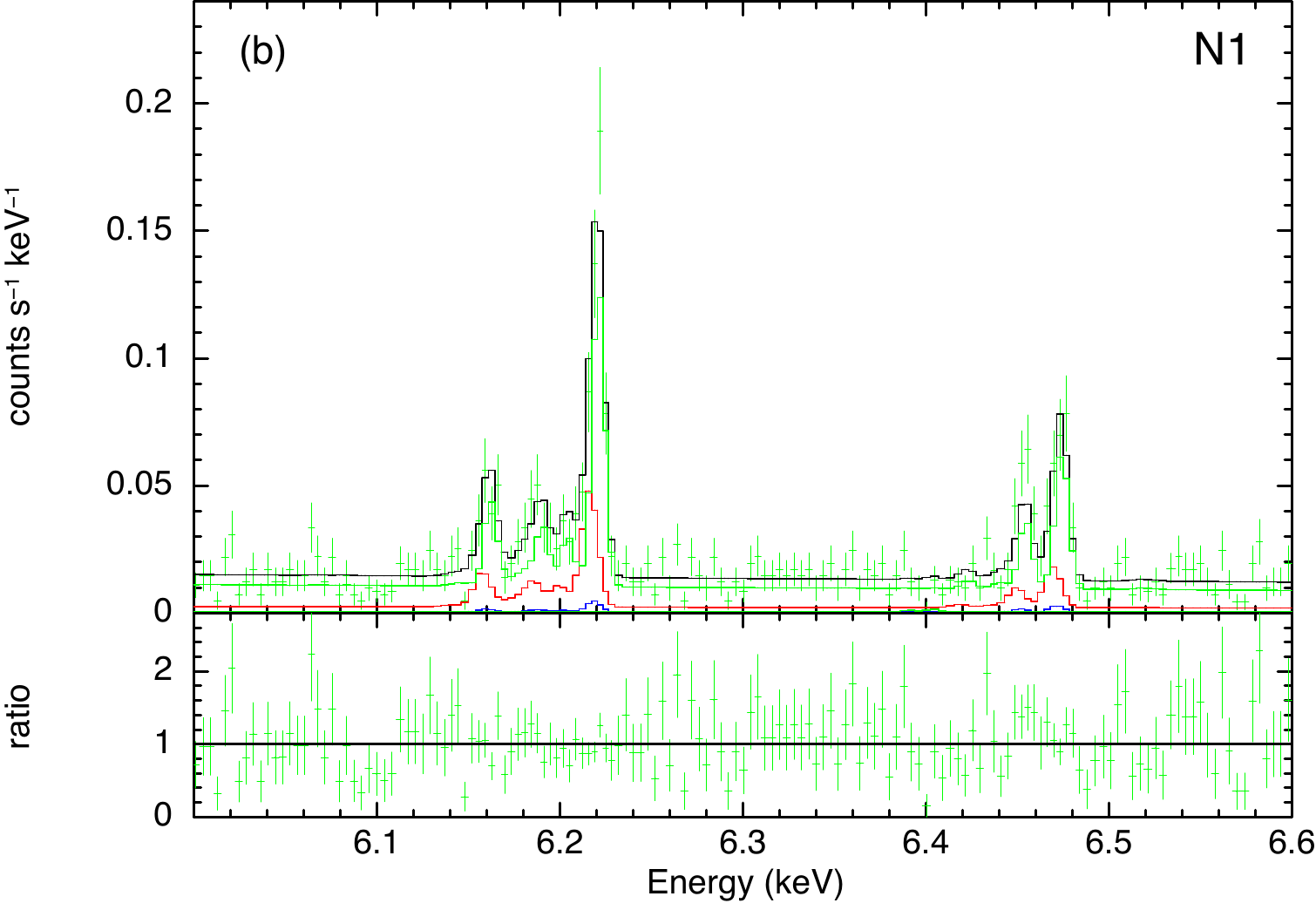}
    \includegraphics[width=0.45\linewidth]{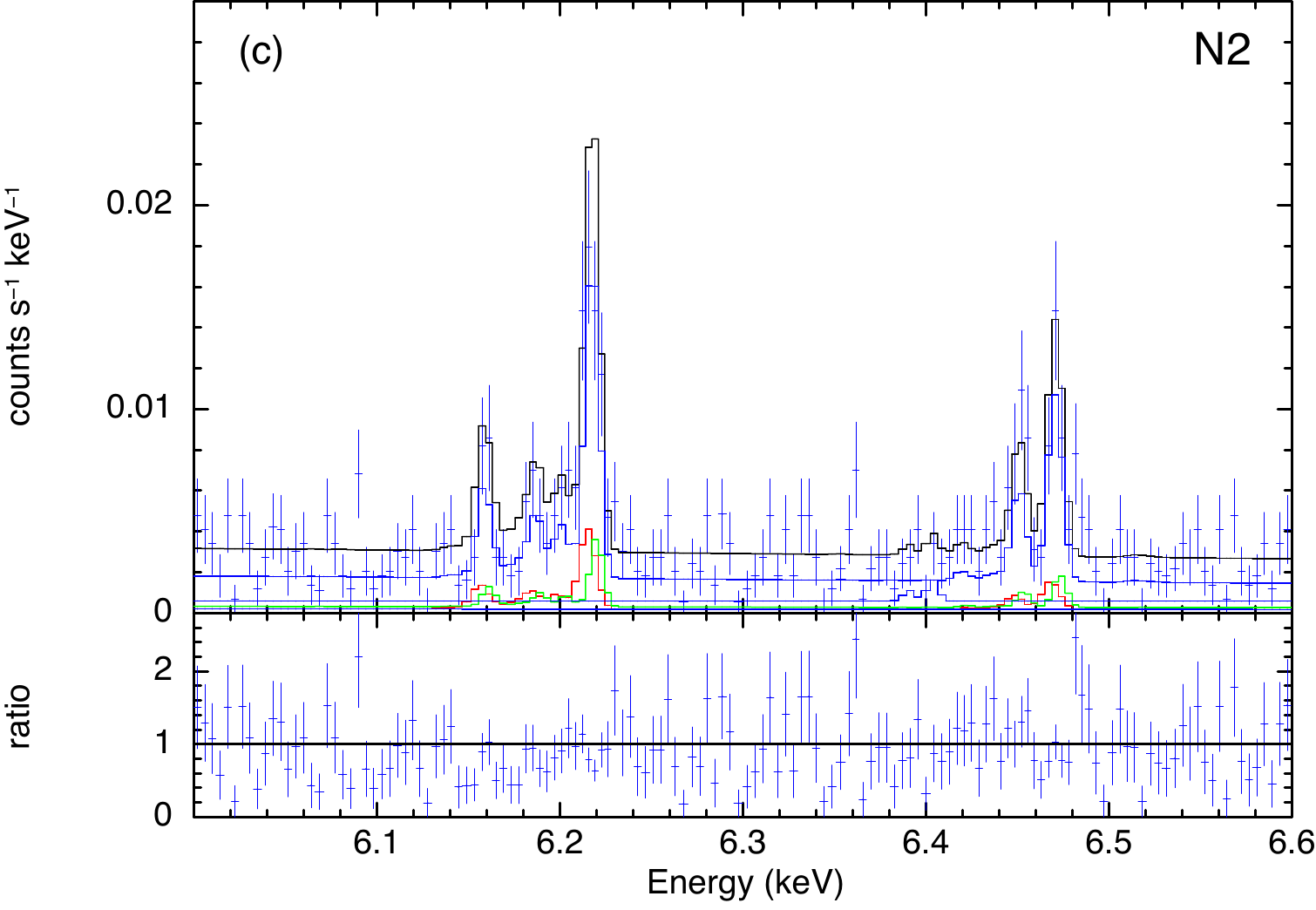}
    \end{center}
    \caption{Same as figure~\ref{fig:spectral_fitting}, but zoomed-in on the iron emission lines. Panels (a), (b), and (c) correspond to the C0, N1, and N2 regions, respectively. In these panels, the contributions from the C0, N1, and N2 regions are shown in red, green, and blue, respectively, while the total model is represented by the black solid line.  {Alt text: Three line graphs showing spectra in units of counts per second per kilo electron volt and their fitted curves. Panels (a), (b), and (c) focus on the 6 to 6.6 kilo electron volt range. }}
    \label{fig:spectral_fitting_iron}
\end{figure*}

Figures~\ref{fig:spectral_fitting} and \ref{fig:spectral_fitting_iron} show the simultaneous fitting of the three regions, taking into account the cross-region PSF effects. Table~\ref{tab:spectral_parameters} summarizes the best-fit parameters with and without the cross-region PSF corrections. Table~\ref{tab:crosstalk} summarizes the fraction of photon counts in the 2--10 keV band and iron emission lines for the pointing regions. From the table, the contribution from emissions outside the region of interest is non-negligible. In the next subsection, we quantitatively evaluate the systematic uncertainties, including this effect.

\begin{table}[htb]
\tbl{Fraction of photons at the 2--10~keV and Fe XXX He$\alpha$ line in the observation regions originating from the annular regions.}{
\centering
\begin{tabular}{lllllll}\hline\hline
     & \multicolumn{3}{c}{2--10~keV band} & \multicolumn{3}{c}{Fe He line} \\ \cline{2-4} \cline{5-7}
     &   A1 &  A2 &  A3  & A1 & A2  & A3  \\ \hline
  C1 & 0.85 & 0.15 & 0.00 & 0.93 & 0.07 & 0.00 \\
  N1 & 0.19 & 0.78 & 0.03 & 0.34 & 0.64 & 0.02 \\
  N2 & 0.18 & 0.13 & 0.69 & 0.20 & 0.12 & 0.68 \\\hline
\end{tabular}}\label{tab:crosstalk}
\end{table}

\begin{table*}[htb]
\tbl{BAPEC model fitting results with and without PSF correction and derived non-thermal pressure fractions.}{
\begin{tabular}{llllllllll}
\hline\hline
Region & $kT$ & $Z$  & $v_{\rm bulk}$ & $\sigma_v$ & norm$^a$ & C-stat/d.o.f. & $\chi^2$/d.o.f.$^b$ & $\alpha$$^c$ & $\alpha_{\rm turb}$$^c$\\ 
&  (keV) & (solar) & ${\rm (km\,s^{-1})}$ & ${\rm (km\,s^{-1})}$ & ${\rm (10^{-2}cm^{-5})}$ & & & & \\
\hline
Center & $6.62^{+0.11}_{-0.11}$ & $0.642^{+0.021}_{-0.024}$ & $13^{+13}_{-7}$ & $148^{+13}_{-9}$ & $4.039^{+0.042}_{-0.037} $ & 7189/7333 & 1490/1539 & $2.1\pm0.3$ &  $2.1\pm0.3$\\
N1 & $7.49^{+0.21}_{-0.21}$ & $0.336^{+0.029}_{-0.023}$ & $-218^{+16}_{-25}$ & $58^{+37}_{-48}$ & $4.584^{+0.067}_{-0.077} $ & & & $1.6\pm0.5$ & $0.3(<0.7)$ \\
N2 & $8.26^{+0.34}_{-0.31}$ & $0.371^{+0.038}_{-0.039}$ & $-93^{+31}_{-20}$ & $94^{+44}_{-50}$ & $2.010^{+0.043}_{-0.039} $ &  & & $0.9\pm0.7$ & $0.7(<1.3)$ \\\hline
Center & $6.38^{+0.13}_{-0.11}$ & $0.610^{+0.024}_{-0.022}$ & $-9^{+12}_{-7}$ & $162^{+10}_{-10}$ & $4.634^{+0.071}_{-0.074}$ & 3653/3634 & 854/803 & & \\
N1 & $7.08^{+0.21}_{-0.2}$ & $0.382^{+0.025}_{-0.022}$ & $-153^{+13}_{-19}$ & $133^{+20}_{-15}$ & $6.002^{+0.100}_{-0.100}$ & 1772/1888 & 391/429 & & \\
N2 & $8.13^{+0.42}_{-0.40}$ & $0.364^{+0.040}_{-0.039}$ & $-114^{+31}_{-16}$ & $110^{+34}_{-42}$ & $2.463^{+0.076}_{-0.072}$ & 1799/1811 & 280/307 & & \\\hline
\end{tabular}}\label{tab:spectral_parameters}
\begin{tabnote}
The top three rows show the results from simultaneous fitting of three pointings with PSF correction applied, while the bottom three rows present the results from individual fits without PSF correction. The uncertainties in the table represent 1$\sigma$ statistical errors. $^a$The normalization (norm) values correspond to the input surface brightness distribution area. $^b$As a reference, the chi-squared values were calculated based on the fit results using spectra rebinned to contain at least 25 counts per bin. $^c$Non-thermal pressure fractions $\alpha$ and $\alpha_{\rm turb}$ are calculated using equations~\ref{eq:fnt} and \ref{eq:fnt_turb}, respectively. $\alpha$ includes both turbulent and bulk velocity contributions, while $\alpha_{\rm turb}$ considers only the turbulent component.
\end{tabnote}
\end{table*}

\subsection{Systematic uncertainties}\label{subsec:systematic}
\begin{figure*}[htb]
    \begin{center}
    \includegraphics[width=0.90\linewidth]{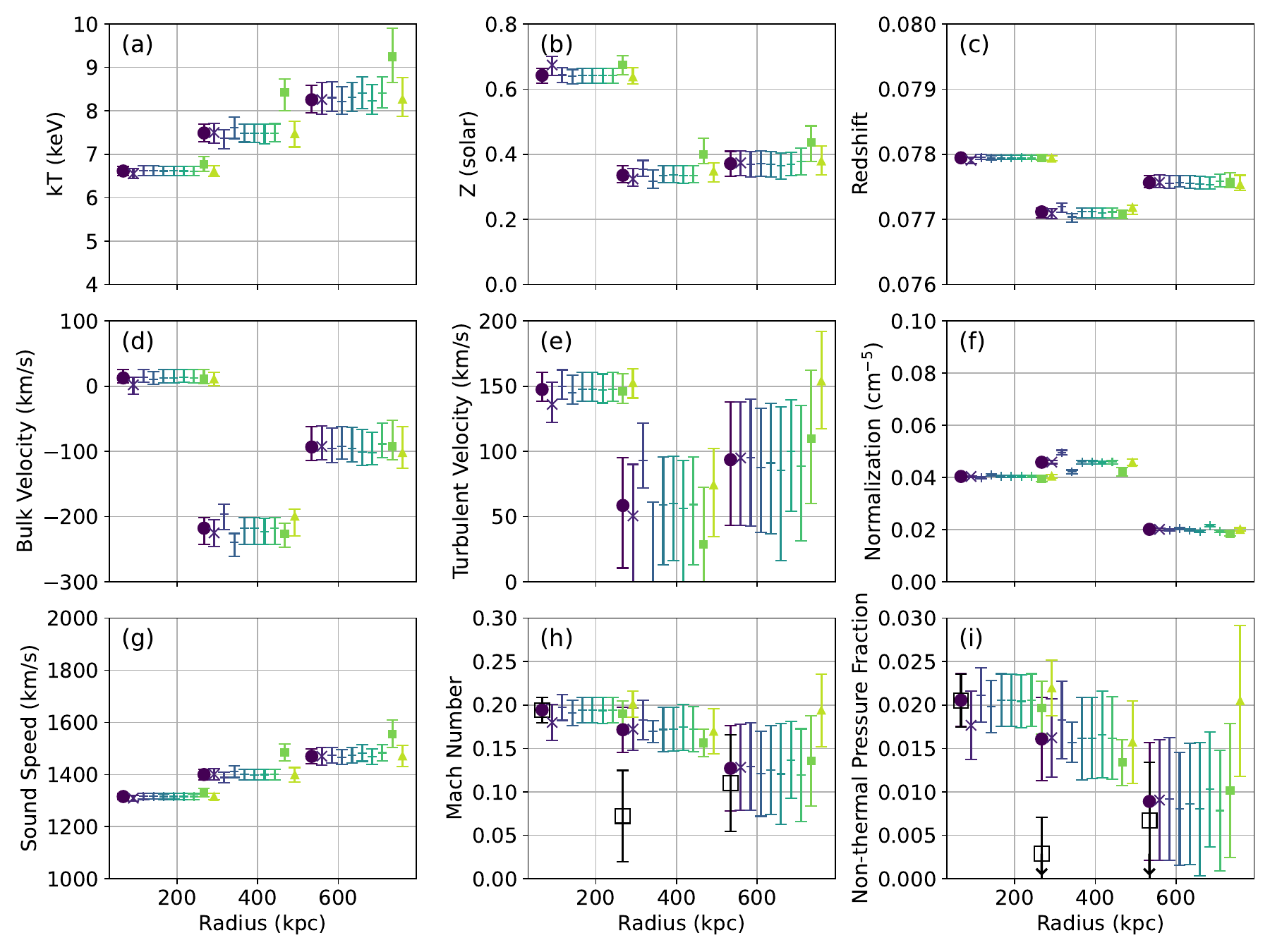}
    \end{center}
    \caption{Radial distribution of the measured properties of the intracluster gas. The nine panels, from top to bottom, show (a) temperature, (b) metal abundance, (c) redshift, (d) line-of-sight bulk velocity, (e) line-of-sight turbulent velocity, (f) normalization, (g) sound speed, (h) effective Mach number (equation~\ref{eq:Meff}), and (i) the fraction of non-thermal pressure (equation~\ref{eq:fnt}). The ten data points, from left to right, correspond to the fit results in the 2--10 keV band (the filled circle); excluding the w-emission lines (the cross); assuming an uncertainty of -30\% and +30\% in the effective area (the plus signs), which are shown in the order of the three combinations of source and spectral regions (A1 to N1, A1 to N2, A2 to N2); the results in the 5.5--7~keV band (the filled square); and the results using no rebinning (the filled triangle). In panels (h) and (i), the Mach number considering only turbulence and the corresponding non-thermal pressure fraction (equation~\ref{eq:fnt_turb}) are indicated by open squares. The x-axis values represent the center of each radial bin (kpc). For each radial bin, the results in the 2--10 keV band represent the standard case and are plotted at the leftmost position, while the other data points are slightly shifted to the right to avoid overlap. {Alt text: Nine line graphs showing data across a horizontal axis ranging from a radius of 0 to 670 kilo parsec.} } 
    \label{fig:radial_distributions}
\end{figure*}

We estimate the systematic errors for the following factors that could affect the gas velocity measurements. i) Resonant scattering, ii) Uncertainty in the telescope's point spread function, iii) Energy range, iv) Spectral binning, v) gain calibration accuracy, and vi) the background model.

i) Resonant scattering: As discussed in \citet{XRISM25}, resonant scattering is expected to be significant in the central region, but it was not detected with a high statistical significance in the current exposure. Therefore, we performed spectral fitting of the three regions again, excluding the w-emission line (the strongest resonance transition in the Fe XXV He$\alpha$ triplet) in the central region. Figure~\ref{fig:radial_distributions} shows that the velocity remains unaffected within the range of statistical errors.

ii) Uncertainties in the telescope's PSF: 
The calibration of the XMA sub-PSF (corresponding to the Resolve pixel size) can have uncertainties of up to 80\% for a point source. However, for an extended source, the PSF is smoothed, reducing these uncertainties. To assess the impact, we tested the effect of a $\pm30$\% systematic uncertainty in the net effective area (summed over pixels). This choice is justified, as a comparison of pixel-to-pixel relative count rates with corresponding ray-tracing predictions shows agreement within 30\% for most pixels.

We applied this $\pm30$\% variation to the effective area for three different combinations of source and spectral regions: (A1 to N1, A1 to N2, A2 to N2). As shown in figure~\ref{fig:radial_distributions}, while the best-fit values fluctuate, the uncertainty ranges overlap, indicating that the systematic effect does not significantly impact the results.

iii) Energy range: To evaluate the impact of the energy range selection, we performed the fit within a narrow band (5.5--7 keV, covering the iron emission line band) instead of the full 2--10 keV range. The results remain consistent within statistical errors.

iv) Spectral binning: Thus far, we have used spectra with binning applied. We also compared the parameters obtained using the original spectral bins without grouping. When using the unbinned spectra, the best-fit value for the turbulent velocity tended to be larger in the N2 region when compared to the rebinned case. This trend could be due to low photon statistics in the N2 region. In any case, the turbulent velocity value was small. To ensure a conservative estimate, we quote the upper limit.

v) Accuracy of the energy scale and instrumental line widths: We reviewed the energy scale calibration reports  \footnote{\url{https://heasarc.gsfc.nasa.gov/docs/xrism/analysis/gainreports/index.html}} for four observations. Correcting the energy scale of the dedicated calibration pixel, which is continuously illuminated, based on the same calibration time intervals as the main array, makes it a witness for the goodness of the intermittent calibration for the main array.  We found the worst energy offset, 0.12 eV, in OBSID 000149000.  Adding in quadrature with the current energy-scale accuracy from 5.4--9.0-keV of $\pm0.3$~eV results in $\pm0.32$~eV, and $\sim$0.3 eV for all three pointings, resulting in a systematic line-of-sight bulk-velocity uncertainty of 15 ${\rm km\,s^{-1}}$ at 6 keV.

The analysis of the calibration pixel in these data sets resulted in a range of FWHM widths from 4.45 to 4.73 eV.  The current in-orbit calibration has a line-spread uncertainty of $\sim0.15$ eV from 6--7 keV.  This error at 6 keV results in a systematic uncertainty in turbulent velocity of  $\pm6\,{\rm km\,s^{-1}}$for turbulence of 50 ${\rm km\,s^{-1}}$ and $\pm2\,{\rm km\,s^{-1}}$for turbulence of 150 ${\rm km\,s^{-1}}$.

As a result, the measured gas turbulence and bulk motions (table~\ref{tab:spectral_parameters}) are largely dominated by statistical errors, except for the bulk velocity in the central region. For the evaluation of effective three-dimensional velocity discussed in the next subsection, the values in the central region are primarily determined by the velocity dispersion. Therefore, the impact of instrumental systematic errors on the velocity measurements across all three regions is expected to be negligible.

vi) Background model: In the central region, the contribution of the background is negligible. However, in the N1 and N2 regions, NXB and CXB contribute approximately 5\% and 2\% for N1, and 23\% and 6\% for N2, respectively, in the 2--10 keV band. To evaluate the impact on the results, we varied the NXB and CXB levels by $\pm$30\%. This test confirmed that there were no statistically significant effects on the measurements of redshift or turbulent velocity. For temperature and metal abundance, we refer readers to Sarkar et al. (in prep.) for further details.

Among the six systematic effects evaluated, only the four that directly influence the spectral fitting results (i--iv) are shown in figure~\ref{fig:radial_distributions}. The remaining two, v) gain calibration and vi) background modeling, were assessed separately. Gain uncertainty was added as a systematic error to the velocity parameters, and variations in the background level were tested but found to have negligible impact on redshift and velocity measurements.

\subsection{Estimation of non-thermal pressure component}
\label{sec:NTestimate}

Based on the simultaneous spectral fitting results of the three regions, we estimate the contribution of non-thermal pressure.

First, we evaluate two velocity components of gas motion -- bulk and turbulent -- which can contribute to non-thermal pressure support and impact hydrostatic mass estimates. The bulk velocity is defined as the line-of-sight velocity relative to the brightest cluster galaxy (BCG), given by  
\begin{equation}
    v_{\rm bulk} = \frac{c (z - z_{\mathrm{BCG}})}{1 + z_{\mathrm{BCG}}},
\end{equation}
where $z$ is the redshift of the gas, and $z_{\mathrm{BCG}}$ is the redshift of the BCG, determined to be 0.0779 based on MUSE observations \citep{XRISM25}. In the central region, the bulk velocity is consistent with zero within the statistical errors. However, in the N1 and N2 regions, significant blueshifts of approximately $-220~\rm{km\,s^{-1}}$ and $-90~\rm{km\,s^{-1}}$, respectively, were observed. 

The turbulent velocity shows variation depending on the region. In the central region, the turbulent velocity is the largest, measured at $148^{+13}_{-9}~\rm{km\,s^{-1}}$. In the intermediate N1 region, the turbulent velocity measurement is an upper limit, $58 (<186)~{\rm km\,s^{-1}}$ ($3\sigma$). In contrast, the N2 region exhibits a marginal detection of turbulent velocity at $94^{+44}_{-50}~\rm{km\,s^{-1}}$.

Next, the sound speed is estimated as  
\begin{equation}
    c_s = \sqrt{\frac{\gamma k_B T}{\mu m_p}},
    \label{eq:cs}
\end{equation}
where $\gamma=5/3$ is the adiabatic index, $k_B$ is the Boltzmann constant, $T$ is the temperature, $\mu$ is the mean molecular weight, and $ m_p$ is the proton mass. Taking into account both turbulent and bulk velocities, the Mach number of the effective three-dimensional velocity \citep{Hitomi18} is defined as
\begin{equation}
    \mathcal{M}_{\rm 3D, eff} = \frac{ \sigma_{v,{\rm eff}} }{c_s} = \frac{\sqrt{3\sigma_v^2 + v_{\rm bulk}^2}}{c_s},
    \label{eq:Meff}
\end{equation}
where $\sigma_v$ is the 1D turbulent velocity dispersion. We note that equation~\ref{eq:Meff} assumes isotropy for the turbulent component, hence the factor of 3, while the bulk velocity is treated as a directional quantity without assuming isotropy. However, it inherently carries some uncertainty due to the unknown orientation of the full 3D velocity field. We address this limitation in section~\ref{subsec:indirect}. 
Using equation~\ref{eq:Meff}, the ratio of non-thermal pressure to the total pressure is estimated as 
\begin{equation}
   \alpha = \frac{P_{\mathrm{NT}}}{P_{\mathrm{tot}}} = \frac{\mathcal{M}_{\rm 3D, eff}^2}{\mathcal{M}_{\rm 3D, eff}^2 + \frac{3}{\gamma}}. \label{eq:fnt}
\end{equation}

The results are summarized in figure~\ref{fig:radial_distributions}. To examine the contribution of turbulent motion alone, we also compute the Mach number defined as $\mathcal{M}_{3D}=\sqrt{3}\sigma_v/c_s$ and the corresponding non-thermal pressure fraction, 
\begin{equation}
\alpha_{\rm turb} = \frac{\mathcal{M}_{\rm 3D}^2}{\mathcal{M}_{\rm 3D}^2 + \frac{3}{\gamma}}.\label{eq:fnt_turb}
\end{equation}
In the core region, the estimated turbulent pressure fraction is $\alpha_{\rm turb} = 2.1\pm0.3$\%, in agreement with the analysis in \citet{XRISM25}. 

The values of the non-thermal pressure fractions estimated from equations~\ref{eq:fnt} and \ref{eq:fnt_turb} are summarized in Table~\ref{tab:spectral_parameters}. At all radial positions, the non-thermal pressure fraction remains small, at $2\%$ or less. The radial distribution shows a trend of decreasing from the central region outward, with $\alpha$ approximately halving from the Center to the N2 region. A more quantitative treatment of the pressure distribution is provided in the next subsection.  

\subsection{Mass modeling with non-thermal pressure}\label{subsec:massmodeling}

Given the estimate of the non-thermal pressure provided in Sect. \ref{sec:NTestimate}, we can attempt to recover the mass profile of the system in the presence of kinetic motions and estimate the hydrostatic mass bias, $b=1-M_{\rm hyd}/M_{\rm tot}$, with $M_{\rm hyd}$ the mass that is recovered under the assumption that the gas in hydrostatic equilibrium. To this aim, we make use of the Python package \texttt{hydromass}\footnote{\url{https://hydromass.readthedocs.io}} \citep{Eckert22}, which allows to reconstruct the mass profiles of galaxy clusters by jointly fitting the surface brightness and spectroscopic temperature profiles. In this framework, the 3D mass profile of the system is described using a parametric mass model $M(<r)$ and combined with the model 3D gas density profile to predict the total pressure,

\begin{equation}
    P_{\rm tot}(r) = P_0 + \int_{r_0}^r \frac{GM(<r)}{r^2}\rho_{\rm gas}(r)\, {\rm d}r
    \label{eq:hse}
\end{equation}

\noindent
with $r_0$ the outermost boundary of the available surface brightness profile, and $P_0=P(r_0)$ the integration constant, that represents the pressure at the outer boundary. In the usual case where non-thermal pressure is neglected, it is assumed that the total pressure described by equation~\ref{eq:hse} can be entirely ascribed to the thermal pressure $P_{T}=\frac{k_b}{\mu m_p}\rho_{\rm gas}T$, such that the model pressure and density can be combined to predict the gas temperature and compare the model to the available spectroscopic temperature data. For the details of the procedure, we refer the reader to \citet{Eckert22}. Here we extend the framework to include measurements of gas velocities, which allow us to constrain departures from hydrostatic equilibrium. Following \citet{Eckert19}, we assume that the residual gas motions detected by XRISM/Resolve are isotropic, such that the spatial velocity term of the Euler equation can be described as an additional pressure term (e.g., \cite{Lau09}),

\begin{equation}
    P_{\rm tot} = P_{T} + P_{NT}
\end{equation}
with 
\begin{equation}
    P_{NT} = \frac{1}{3}\rho_{\rm gas}\sigma_{v,3D}^2. 
    \label{eq:Pnt3D}
\end{equation}
Here, $\sigma_{v,3D}$ represents the 3D turbulent velocity dispersion.

Considering a model for $P_{NT}(r)$, the profile of $\sigma_{v,3D}$ can be calculated within the model. To allow for a proper comparison to the data, we remark that the estimated $\sigma_{v,{\rm eff}}$ described in equation~\ref{eq:Meff} is averaged over the velocity structure along the line of sight, weighted by the local emission measure, $EM_{3D}$ \citep{Truong2024},

\begin{equation}
    \sigma_{v,{\rm eff}} = \sqrt{3\sigma_{v}^2 + v_{\rm bulk}^2} = \left[ \frac{\int EM_{3D} \sigma_{v,3D}^2 \, {\rm d}\ell}{\int EM_{3D} \, {\rm d}\ell} \right]^{1/2}.
\end{equation}

For $P_{NT}(r)$, we use the parametric function proposed by \citet{Angelinelli20}, which was found to provide a good fit to the non-thermal pressure profiles of simulated clusters in the Itasca simulation \citep{Vazza2017}. The radial dependence of non-thermal pressure ratio in equation~\ref{eq:fnt} is described as a power law with a constant floor,

\begin{equation}
    \alpha_{NT}(r) = \frac{P_{NT}}{P_{\rm tot}} = a_0\left( \frac{r}{R_{200,m}}\right)^{a_1} + a_2
    \label{eq:angelinelli}
\end{equation}

The parameters $a_0$, $a_1$, and $a_2$ govern the radial evolution of the non-thermal pressure fraction, whereas $R_{200,m}$ represents the radius within which the mean density is 200 times the mean density of the Universe. 
For more details on the implementation of the non-thermal pressure model, we refer to Chappuis et al. (in prep.).

We complement the Resolve data with publicly available measurements of the gas density and temperature profiles of the system from XMM-Newton data obtained in the framework of the XMM Cluster Outskirts Project (X-COP; \cite{XCOP,Ghirardini2019})\footnote{\url{https://dominiqueeckert.wixsite.com/xcop/a2029}}. These radial profiles were extracted from a set of 7 individual XMM-Newton pointings and cover the radial range from the core out to $R_{200}\sim 2.1$ Mpc. We note that these profiles are radially averaged, whereas the Resolve data were obtained over a single arm covering a fraction of the azimuth; thus, we make the assumption that the velocity structure presented here is representative of the full azimuth.

We run \texttt{hydromass} on a dataset made of the Resolve $\sigma_{v,{\rm eff}}$ measurements, the X-COP spectroscopic temperature, and the X-COP surface brightness profile. We model the mass profile $M(<r)$ as a Navarro-Frenk-White (NFW) profile \citep{nfw96,nfw97}, with the concentration $c_{200}$ and the overdensity radius $R_{200}$ as free parameters of the model. The full model has six free parameters: the NFW parameters $c_{200}$ and $R_{200}$, the integration constant $P_0$, and the parameters describing $\alpha_{NT}(r)$, i.e. $a_0$, $a_1$ and $a_2$ (Equation~\ref{eq:angelinelli}). We set uniform priors on both NFW parameters, with an allowed range of [1-20] for $c_{200}$ and [500-4000] kpc for $R_{200c}$. To set the prior on $P_0$, we fit the pressure profile of the system with a functional form and evaluate the model pressure at the edge of the available data. We then set a uniform prior on $P_0$ centered on the estimated value and encompassing uncertainties of $\pm50\%$. Finally, we set uniform priors on the three parameters of the \citet{Angelinelli20} profile: $a_0\sim\mathcal{U}(-0.5, 2.0)$, $a_1\sim\mathcal{U}(-1, 2)$, and $\log a_0\sim\mathcal{U}(-6, 0)$. The model is optimized using the No U-Turn Sampler (NUTS) as implemented within the \texttt{PyMC} package \citep{pymc}. 

\begin{figure*}
    \begin{center}
    \includegraphics[width=0.5\linewidth]{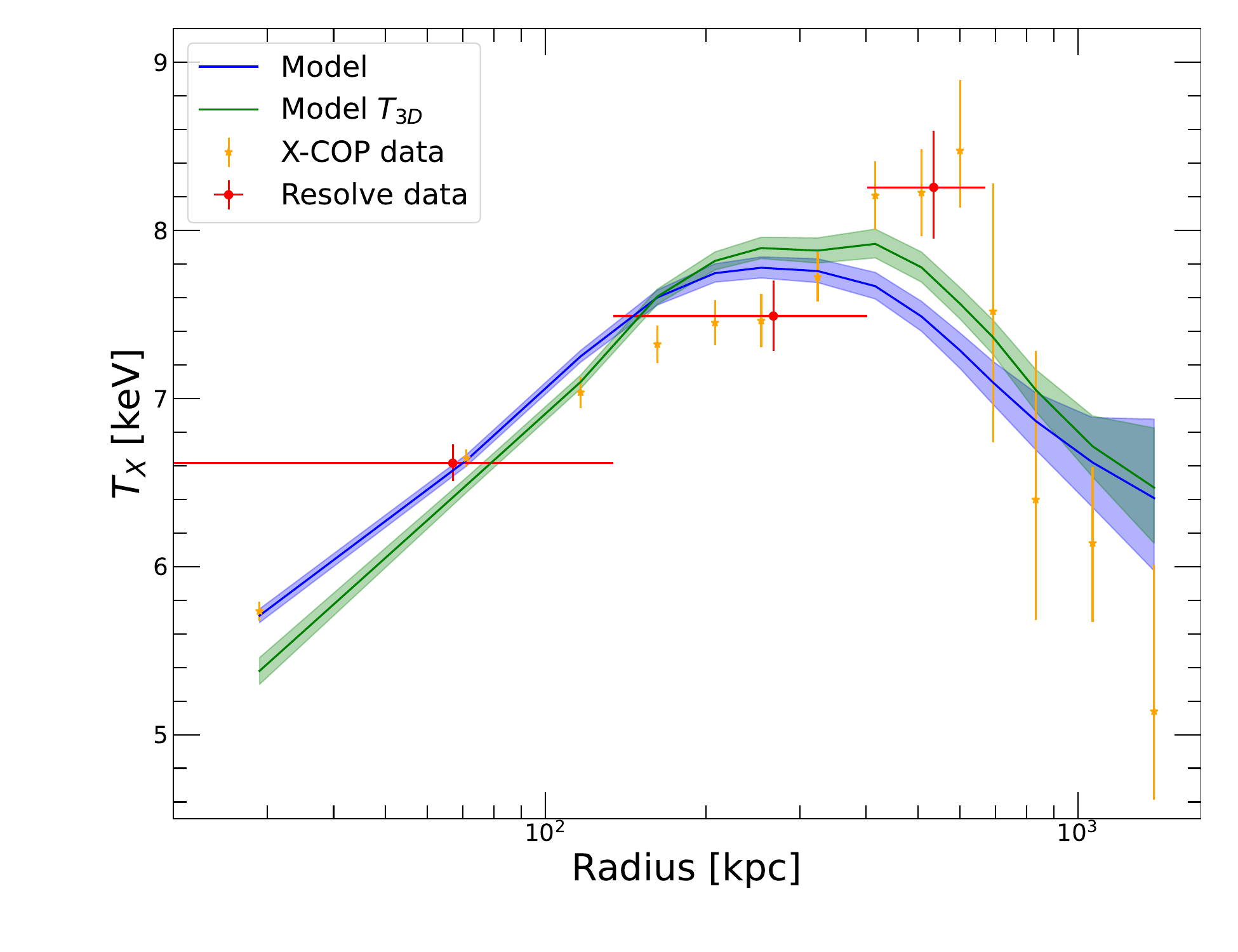}\includegraphics[width=0.5\linewidth]{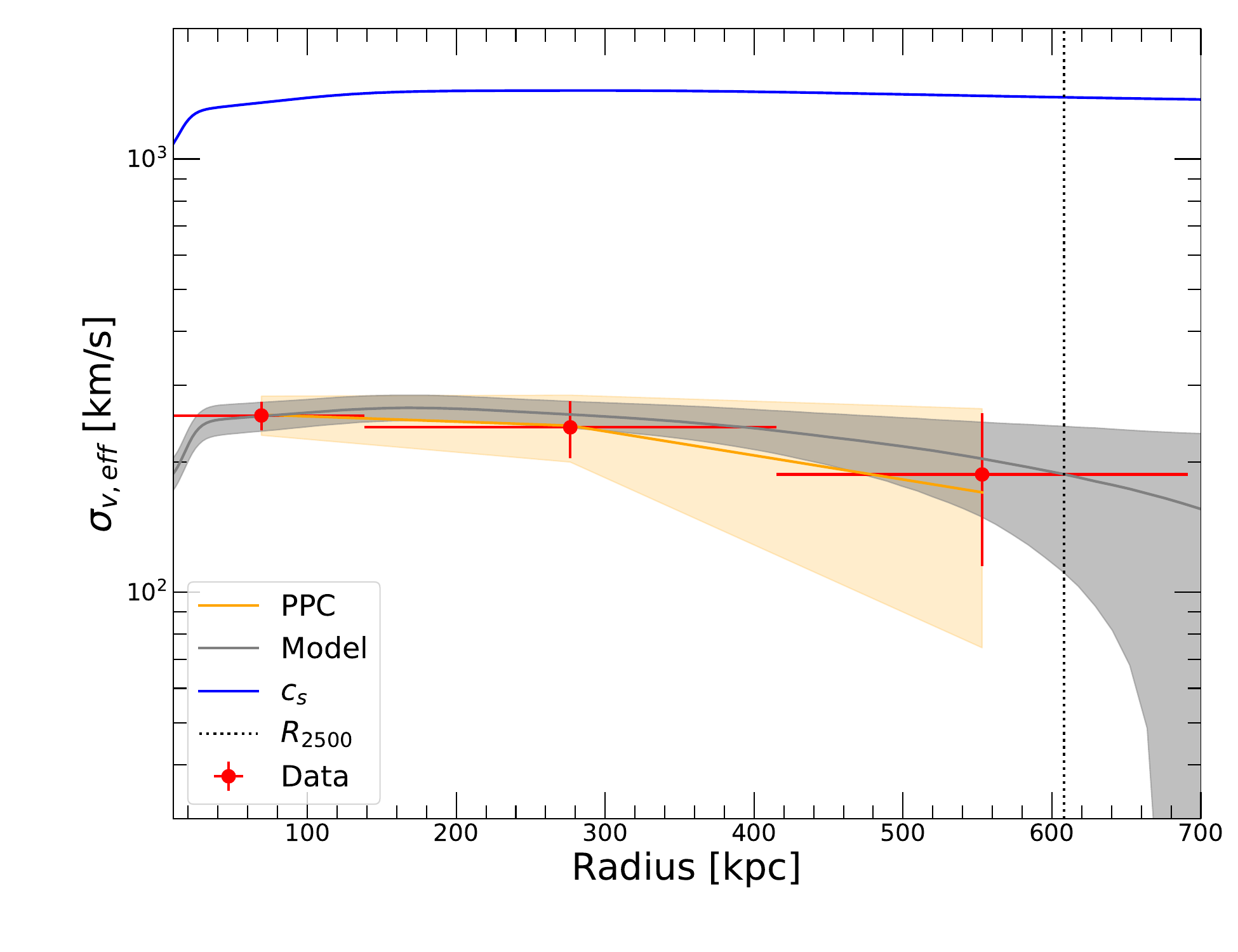}
    \end{center}
    \caption{Result of the hydrostatic NFW mass reconstruction. The left-hand panel shows the model spectroscopic temperature profile (blue curve and shaded area) and its deprojected 3D counterpart (green). The XMM-Newton X-COP data points are shown in orange, whereas the red points provide the temperature measurements obtained by XRISM/Resolve (see Table \ref{tab:spectral_parameters}). The right-hand panel shows the fit to the effective velocity dispersion $\sigma_{v,{\rm eff}}$ (Equation~\ref{eq:Meff}, red data points) with the \citet{Angelinelli20} parametrization (gray curve and shaded area). The orange curve shows posterior predictive checks (PPC) generated from the best-fit model, which show that the model provides a good representation of the data. For comparison, the blue curve shows the sounds speed in the medium (Equation~\ref{eq:cs}). The approximate location of $R_{2500}$ is highlighted by the vertical dotted line. {Alt text: Two line graphs showing results from mass modeling. The horizontal axis in the left panel ranges from a radius of 2 to 2000 kilo parsecs, while in the right panel it ranges from 0 to 700 kilo parsecs.}}
    \label{fig:hydromass_fit}
\end{figure*}

The results of the fitting procedure are shown in figure~\ref{fig:hydromass_fit}, where the left-hand panel displays the Resolve and X-COP spectroscopic temperature profiles and the fitted temperature model. We note the excellent agreement between the two datasets, which implies that our assumption that the Resolve results are representative of the full azimuth is probably justified. The right-hand panel of figure~\ref{fig:hydromass_fit} displays the effective velocity dispersion profile (Equation~\ref{eq:Meff}) as well as the best-fit model using the \citet{Angelinelli20} parametrization (Equation~\ref{eq:angelinelli}). Posterior predictive checks (PPC) generated from the best-fit model agree well with the $\sigma_{v,{\rm eff}}$ data, which shows that the model provides a good representation of the data at hand. For comparison, we also show the sound speed profile in the same figure, which allows to readily convert the model velocity dispersion into a turbulent Mach number of $\sim0.2$ (see Sect. \ref{sec:NTestimate}). 

\begin{table*}
    \caption{Results of hydrostatic mass reconstruction in the presence of non-thermal pressure. The listed parameters are the masses (in unit of $10^{14} M_\odot$) and overdensity radii (in kpc) within overdensities of 2500 and 200 times the critical density, the NFW concentration $c_{200}=R_{200}/r_s$, where $r_s$ is the scale radius of the NFW profile. Additionally, the table includes the parameters of the \citet{Angelinelli20} profile (Equation~\ref{eq:angelinelli}).}
    \centering
    \begin{tabular}{cccccccc}
    \hline \hline
       $M_{2500}$ & $R_{2500}$ & $M_{200}$ & $R_{200}$ & $c_{200}$ & $a_0$ & $a_1$ & $a_2$ \\      
       \hline
       $3.4\pm0.1$ & $608\pm4$ & $9.5\pm0.3$ & $1979\pm21$ & $5.9\pm0.2$ & $-0.13\pm0.12$ & $1.6\pm0.5$ & $0.021\pm0.004$ \\
       \hline
    \end{tabular}
    \label{tab:mass_reconstruction}
\end{table*}

The best-fit NFW model has a total mass $M_{200}=(9.5\pm0.3)\times10^{14}M_\odot$, which is only 2\% larger than the hydrostatic reconstruction extracted from the same data \citep{Eckert22}. The best-fitting parameters for the NFW model and the NT pressure ratio are shown in table~\ref{tab:mass_reconstruction}. We can see that the floor parameter of the \citet{Angelinelli20} model, $a_2$, is very well constrained at a value of $\sim2\%$, whereas the radial trend of $\alpha_{NT}(r)$ is relatively poorly constrained. Technically, the fitted profile of $\alpha_{NT}$ becomes negative beyond $\sim1,500$ kpc, which leads to an unphysical velocity dispersion; in such a case, our code sets the predicted value of $\sigma_{v,3D}$ to zero. Constraints on the velocity field at larger radii and throughout a more representative region of the cluster are needed to better determine the radial trend of $\alpha_{NT}(r)$.

\begin{figure}
    \begin{center}
    \includegraphics[width=\linewidth]{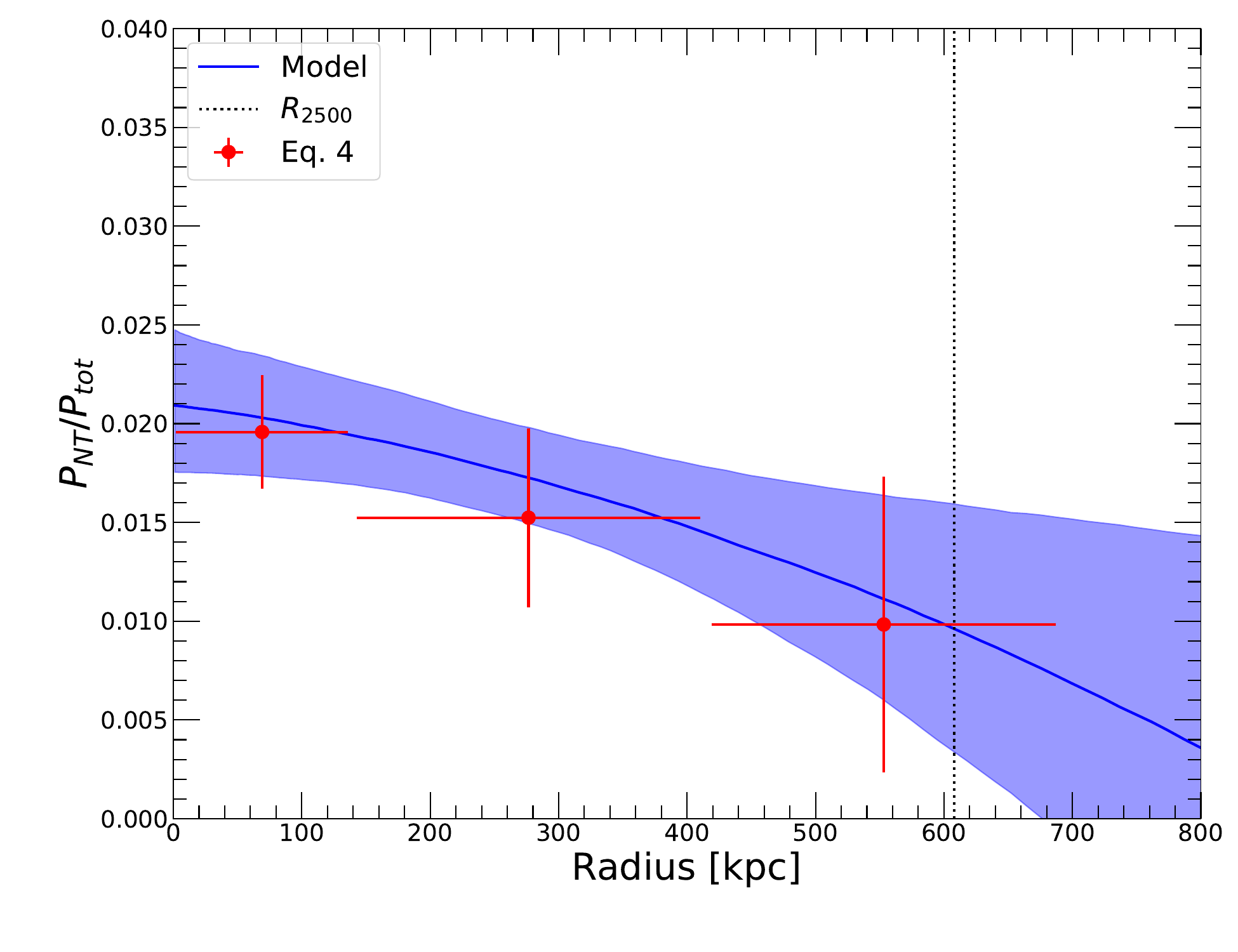}
    \end{center}
    \caption{Non-thermal pressure ratio as a function of cluster-centric radius inferred from the best-fit hydrostatic model (blue line and shaded area). The red data points show the non-thermal pressure ratio obtained directly from the Resolve data points through equation~\ref{eq:fnt}. The dashed vertical line shows the location of $R_{2500}$. {Alt text: Line graph showing the pressure ratio. The horizontal axis ranges from a radius of 0 to 800 kilo parsec, and the vertical axis ranges from 0 to 0.04.}}
    \label{fig:alphaNT}
\end{figure}

These results show that the hydrostatic bias $b=1-M_{\rm hyd}/M_{\rm tot}$ is very low, of the order of 0.02 across the radial range of interest. The model $\alpha_{NT}$ profile is shown in figure~\ref{fig:alphaNT}, together with the non-thermal pressure estimates obtained by converting the measured velocities into a non-thermal pressure ratio through equation~\ref{eq:fnt}. We find a good agreement between the estimate obtained with equation~\ref{eq:fnt} and the posterior distribution of the model. 
We note the slight decreasing trend in the radial profile of the non-thermal pressure ratio, although the resulting distribution is consistent with a flat profile at $\sim1\sigma$. Conversely, numerical simulations usually predict an increasing trend of $\alpha_{NT}(r)$ \citep{Lau09,Nelson14,Biffi16,Vazza2017,Angelinelli20}. 

\section{Discussion}\label{sec:discussion}
We derived new constraints on the gas motions outside the core of the galaxy cluster Abell 2029 using Resolve's high-resolution spectroscopic data, placing firm limits on the non-thermal pressure and estimating the hydrostatic mass bias $b$. To verify the reliability of this result, we applied two independent methods: Section~\ref{subsec:massmodeling} presents a forward modeling analysis combining Resolve and XMM-Newton data, while section~\ref{subsec:massbias} evaluates density and pressure profiles based on a deprojection analysis of Resolve data. By comparing these approaches, we aim to confirm the robustness of the mass bias estimation and discuss the implications of our findings for the cluster's gas dynamics and mass estimates.

\subsection{Non-thermal pressure and hydrostatic mass bias}\label{subsec:massbias}
A direct estimate of the hydrostatic mass bias can be obtained through an evaluation of the contribution of the non-thermal component to the total pressure, under the assumption that total ``true'' mass is the one associated to the gravitational potential in equilibrium with the gradient of the total (i.e. thermal plus non-thermal) gas pressure divided by the gas density. Following Equation~6 (with Eqs.~4 and 3) in \citet{Ettori22}, the hydrostatic mass bias $b$ can be written as
\begin{equation}
b(r) =  1 - \frac{M_{\rm hyd}(<r)}{M_{\rm tot}(<r)} = 
\frac{\alpha(r) +A(r)}{1+A(r)},
\end{equation}
where $\alpha = P_{\rm NT}/P_{\rm tot}$ and $A = P_{\rm tot} \, d\alpha/dr \, (d P_{\rm T} / dr)^{-1}$.
Alternatively, using the hydrostatic equilibrium equation for the thermal and total pressure separately allows us to write 
\begin{equation}
\frac{M_{\rm hyd}}{M_{\rm tot}} = \frac{1}{1+ (d P_{\rm NT}/dr) \, (d P_{\rm T}/dr)^{-1}},
\end{equation}
which can be translated into another estimate of the hydrostatic mass bias, $b_2 \approx (d P_{\rm NT}/d P_{\rm T}) / (1 + d P_{\rm NT}/d P_{\rm T})$. 
To make these calculations using only the 3 points available to the Resolve analysis, 
we have estimated the gas density both from the onion-peeling deprojection of the values of the normalization of the {\tt bapec} model (see Table~\ref{tab:spectral_parameters}) and by assuming a universal pressure profile (e.g., \cite{Arnaud10}) and dividing it by the values of the spectroscopic temperatures from the best-fit 1T model. Then, both $A$ and $d P_{\rm NT}/d P_{\rm T}$ can be evaluated by estimating the slopes (and relative errors) of the linear fit of the relations $\alpha - r$, $P_{\rm T} - r$, and $P_{\rm NT} - P_{\rm T}$, respectively.
We consistently obtain $A$ in the range $(0.7 - 2.8) \times 10^{-5}$, which provides a first direct measurement of a quantity previously known only from hydrodynamic simulations (see  e.g., Figure~10 in \cite{Angelinelli20}), and implies values of $b(r)$ between 2.1 ($\pm$0.3) per cent in the central bin and 0.9 ($\pm$0.7) in the outermost annulus. With the second approach, we estimate over the region a $b_2$ of about 2.2 (with a statistical error of 0.4) per cent.

\subsection{Comparison with indirect measurements} \label{subsec:indirect}
The non-thermal pressure contribution to the total pressure varies between $0.9\pm0.7$ and $2.1\pm0.3$\% within $R_{2500}$ (Figure~\ref{fig:radial_distributions}). It is interesting to compare these values with earlier estimates based on indirect probes of velocities of gas motions in the ICM. Recently, \citet{Heinrich24} analyzed X-ray surface brightness fluctuations within $R_{2500}$ in a sample of bright galaxy clusters observed with Chandra, measuring power spectra of density fluctuations and converting them to velocity power spectra assuming that both are proportional to each other with the proportionality coefficient calibrated for clusters in different dynamical states \citep{Zhuravleva23}. This fluctuations analysis obtained the non-thermal pressure contribution between 0.3--2.1\% within the $R_{2500}$ region in A2029, accounting for statistical errors and uncertainties related to the largest contributing scales. Within the uncertainties, the indirect method is in excellent agreement with the direct XRISM Resolve measurements. The same indirect method obtained upper limits on the weighted-average $P_{\rm NT}/P_{\rm tot}$ of $6\pm 2$ and $4\pm2$ \% within $R_{\rm 2500}$ in relaxed and intermediate clusters, respectively, placing A2029 within this range.

These results confirm that the independent methods provide consistent estimates, with the hydrostatic mass bias in the $R_{2500}$ region being as low as 2\%. This suggests that the contribution of non-thermal pressure is minimal and has a negligible impact on cluster mass estimates. However, our estimate assumes isotropic gas motions, while cosmological simulations predict that directional dependence may not be negligible, potentially leading to an underestimation of velocity dispersion when observations are limited to a single azimuthal direction \citep{Ota18}. In addition, simulations of a Perseus-like sloshing core by \citet{ZuHone18}  indicate that assuming isotropic gas motions may result in an underestimation of the total gas velocity. Furthermore, \citet{Lau13} found that random and rotational motions have a greater effect on mass estimates than bulk (streaming) motion. These insights highlight the importance of multi-directional observations to better account for the gas velocity field and improve mass estimates.

Previous studies suggest that gas motions induced by sloshing in the core of A2029 are likely occurring mostly in the plane of the sky, with a viewing angle estimated to be less than $30\degree$ based on Chandra X-ray imaging \citep{Paterno-Mahler13}. Based on the pressure balance between the radio lobes and the ICM ram pressure, they estimate the gas velocity to be approximately $150-300~{\rm km\,s^{-1}}$. Due to projection effects, a viewing angle of $30\degree$ can lead to an underestimation of the inferred gas velocity by a factor of $\sim2$ \citep{ZuHone18}. If the true three-dimensional velocity in our observation is underestimated by a similar factor, the corresponding non-thermal pressure fraction could reach up to $\sim5$\%. Nevertheless, this still indicates a low level of non-thermal pressure support in the core region. Given the uncertainties associated with the 3D structure of gas motions, we plan to perform further analysis by dividing the central region into smaller subregions to search for velocity gradients, and by comparing our results in more detail with tailored numerical simulations of sloshing in A2029-like clusters. However, such efforts are beyond the scope of the present paper.

\section{Conclusions}
In this study, we conducted a detailed investigation of the gas dynamics and non-thermal pressure in the galaxy cluster A2029 using high-resolution spectroscopic data from XRISM Resolve. Resolve’s unprecedented spectral resolution and precise velocity measurements enabled us to directly assess the role of gas motions in hydrostatic mass estimates, a crucial factor in evaluating mass bias. Our analysis yielded the following key findings:
\begin{itemize}
\item The observed bulk and turbulent gas velocities were generally small. In the center, the turbulent velocity reached $150\,{\rm km\,s^{-1}}$ and decreased with radius. Blueshifts of approximately $220\,{\rm km\,s^{-1}}$ and $90\,{\rm km\,s^{-1}}$ were detected in the two northern regions, indicating significant bulk motions outside the core. Both turbulent and bulk motions can contribute to deviations from hydrostatic equilibrium.

\item In the outer regions up to $R_{2500}$, the non-thermal pressure component remained small, below 2\%, and decreased with radius. This trend differs from many simulations predicting an increase at large radii. Our direct velocity measurements yielded a low hydrostatic mass bias of only 2\% across the observed radial range.

\item Our results show good agreement with indirect methods, such as surface brightness fluctuation analyses, further supporting the robustness of our results.
\end{itemize}

These findings indicate that within the scale radius $R_{2500}$, non-thermal pressure plays a minimal role in mass estimates based on hydrostatic equilibrium in relaxed clusters like A2029. This highlights the XRISM Resolve's capability to provide accurate constraints on cluster mass, enhancing its potential as cosmological probe.

However, several open questions remain. Future observations across multiple azimuthal directions are needed to verify the assumption of isotropic gas motions. Additionally, a larger sample of galaxy clusters will be essential to establish statistically significant constraints on non-thermal pressure profiles and hydrostatic mass bias in different cluster environments.

\bibliographystyle{pasj}
\bibliography{ref.bib}

\begin{ack}
We gratefully acknowledge the hard work over many years of all of the engineers and scientists who made the XRISM mission possible. Part of this work was performed under the auspices of the U.S. Department of Energy by Lawrence Livermore National Laboratory under Contract DE-AC52-07NA27344. The material is based upon work supported by NASA under award numbers 80GSFC21M0002 and 80GSFC24M0006. This work was supported by JSPS KAKENHI grant numbers JP23H00121, JP22H00158, JP22H01268, JP22K03624, JP23H04899, JP21K13963, JP24K00638, JP24K17105, JP21K13958, JP21H01095, JP23K20850, JP24H00253, JP21K03615, JP24K00677, JP20K14491, JP23H00151, JP19K21884, JP20H01947, JP20KK0071, JP23K20239, JP24K00672, JP24K17104, JP24K17093, JP20K04009, JP21H04493, JP20H01946, JP23K13154, JP19K14762, JP20H05857, and JP23K03459. This work was supported by NASA grant numbers 80NSSC20K0733, 80NSSC18K0978, 80NSSC20K0883, 80NSSC20K0737, 80NSSC24K0678, 80NSSC18K1684, 80NSSC23K0650, and 80NNSC22K1922. LC acknowledges support from NSF award 2205918. CD acknowledges support from STFC through grant ST/T000244/1. LG acknowledges financial support from Canadian Space Agency grant 18XARMSTMA. NO acknowledges partial support by the Organization for the Promotion of Gender Equality at Nara Women's University. MS acknowledges the support by the RIKEN Pioneering Project Evolution of Matter in the Universe (r-EMU) and Rikkyo University Special Fund for Research (Rikkyo SFR). AT and the present research are in part supported by the Kagoshima University postdoctoral research program (KU-DREAM). SY acknowledges support by the RIKEN SPDR Program. TY acknowledges support by NASA under award number 80GSFC24M0006. IZ acknowledges partial support from the Alfred P. Sloan Foundation through the Sloan Research Fellowship. SE acknowledges the financial contribution from the contracts Prin-MUR 2022 supported by Next Generation EU (M4.C2.1.1, n.20227RNLY3 {\it The concordance cosmological model: stress-tests with galaxy clusters}), ASI-INAF Athena 2019-27-HH.0, ``Attivit\`a di Studio per la comunit\`a scientifica di Astrofisica delle Alte Energie e Fisica Astroparticellare'' (Accordo Attuativo ASI-INAF n. 2017-14-H.0), and from the European Union’s Horizon 2020 Programme under the AHEAD2020 project (grant agreement n. 871158). LL acknowledges the financial contribution from the INAF grant 1.05.12.04.01. This work was supported by the JSPS Core-to-Core Program, JPJSCCA20220002. The material is based on work supported by the Strategic Research Center of Saitama University.
\end{ack}

\end{document}